\documentclass[prd,superscriptaddress,twocolumn,preprintnumbers,%
  amsmath,amssymb]{revtex4-1}
  
\usepackage[breaklinks,colorlinks=true]{hyperref}
\usepackage{mathtools}
\usepackage{color}
\usepackage{slashed}
\usepackage{amsmath,bm}
\usepackage[normalem]{ulem}
\allowdisplaybreaks

\newcommand{\cout}[1]{\if 0 #1 \fi}

\begin{document}
\title{Euler-Heisenberg Lagrangian under an axial gauge field}
\author{Patrick Copinger}
\email{copinger0@gate.sinica.edu.tw}
\affiliation{Institute of Physics, Academia Sinica, Taipei 11529, Taiwan}
\author{Koichi Hattori}
\email{koichi.hattori@zju.edu.cn}
\affiliation{Zhejiang Institute of Modern Physics, Department of Physics, Zhejiang University, Hangzhou, Zhejiang 310027, China}
\affiliation{Research Center for Nuclear Physics, Osaka University, 
10-1 Mihogaoka, Ibaraki, Osaka 567-0047, Japan}
\author{Di-Lun Yang}
\email{dlyang@gate.sinica.edu.tw}
\affiliation{Institute of Physics, Academia Sinica, Taipei 11529, Taiwan}


\begin{abstract}
Augmentations to the Euler-Heisenberg Lagrangian (QED one-loop effective action in homogeneous electromagnetic fields) under a constant background axial gauge are examined. Two special configurations admit an exact eigendecomposition, and hence effective action as a spectral sum, of the augmented Dirac operator: one with a magnetic field with chiral chemical potential, and the other with an electric field with spatial axial gauge, which resembles an emergent vorticity. An enhancement to Schwinger pair production is found for the latter, which is more fully analyzed using the worldline instanton formalism. There it is found the overall enhancement is due to the spatial axial gauge serving as a negative mass shift. Finally, we remark on the exactly solvable massless case for arbitrary electromagnetic and axial gauges.
\end{abstract}

\maketitle

\section{Introduction}

The Euler-Heisenberg Lagrangian~\cite{Heisenberg:1935qt,Weisskopf:1936hya} has proved to be a spectacularly successful theoretical tool with which to study a wealth of physical phenomena in quantum electrodynamics (QED) including but not limited to the following: Schwinger pair production~\cite{sauter,Heisenberg:1935qt,PhysRev.82.664},  light-by-light scattering~\cite{Euler:1935zz,*PhysRev.83.776}, and charge renormalization~\cite{PhysRev.93.615,*PhysRev.94.1362}; see~\cite{Dunne:2004nc,Fedotov:2022ely} for reviews.  The Lagrangian has been extended to inhomogeneous fields~\cite{landau1982quantum,*narozhnyi1970simplest,*DUNNE1998322,PhysRevD.58.105022}, to include a chemical potential~\cite{PhysRevD.42.2881}, to encompass higher-order loops~\cite{ritus1975lagrangian,*ritus1977connection,*Ritus:1998jm,Gies:2016yaa}, and to study under its derivative expansion in~\cite{doi:10.1063/1.533037}, and it remains an active research thread thanks to its all-orders in background coupling construction, imparting a wealth of physics in a compact formula. Even so, studies on the Euler-Heisenberg Lagrangian under a background $CP$-violating axial gauge with electromagnetic field are few despite its importance.  We study a background axial gauge field as an effective theory; let us highlight selected physically significant phenomenological occurrences both for time-like and space-like axial gauge fields.

The temporal component of an axial gauge field could be regarded as an axial chemical potential or the time derivative of an axion-like particle, which plays an important role in pseudoscalar inflation of early Universe when coupled to the (hyper) gauge fields via an effective Chern-Simons term. In such a scenario, backreaction from the Schwinger effect could modify the dynamics of axion inflatons and the corresponding physics in cosmology~\cite{Domcke:2018eki,*Domcke:2019mnd,*Sobol:2019xls} (see also~\cite{Kamada:2022nyt} for a brief review and more references therein). There have been some related studies for the Schwinger effect with an axial chemical potential. E.g., in~\cite{Domcke:2021fee}, it is found that a nonzero $A^0_5$ could assist the pair production. In contrast, the influence from the spatial component of $A^{\mu}_5$ on the Schwinger effect is rarely studied due to the lack of phenomenological applications. 

However, such a background axial gauge field with the non-vanishing spatial component may be generated from a mechanical strain in the effective theory of Dirac or Weyl semimetals~\cite{Zubkov:2015cba,*PhysRevLett.115.177202,*PhysRevX.6.041046,*PhysRevX.6.041021}. Moreover, the spatial component of a background axial gauge field may be realized as a vortical field coming from rotation, while the temporal component is associated with the fluid helicity~\cite{PhysRevD.103.125003}. In phenomenology, the vortical field (or vorticity for brevity) and magnetic field could lead to similar phenomena such as spin polarization or anomalous currents in chiral matter, known as the chiral magnetic effect~\cite{Vilenkin:1980fu,*Nielsen:1983rb,*Alekseev:1998ds,*Fukushima:2008xe} and chiral vortical effect~\cite{Vilenkin:1979ui,*Son:2009tf,*Landsteiner:2011cp}, even though they are intrinsically different objects. It is thus intriguing to further explore how a background axial gauge field as vorticity may affect the Schwinger effect as opposed to its counterpart with magnetic fields that has been widely studied in literature~\cite{nikishov1970pair,*bunkin1970possibility,*PhysRevD.73.065020,Dunne:2004nc}.

To treat the augmentation to the Euler-Heisenberg Lagrangian we examine in addition to a classical electromagnetic gauge field in (3+1)-dimensions, a classical axial gauge field. External field QED arises from a microscopic treatment wherein the dynamical gauge field is treated classically~\cite{Gies:2016yaa}; for an axial gauge background this treatment is marred. Yet, from a phenomenological standpoint, i.e., by mimicking rotation and torsion and or quantifying the strengths of the chiral magnetic effect or pseudoscalar inflation, there is much value in studying a QED inspired model with both electromagnetic and axial gauge minimal background couplings, which we accomplish here. However, neither electromagnetic nor axial gauge dynamical degrees of freedom are treated here.
Accordingly, QED with an axial gauge field later referred to in this manuscript should always be regraded as an effective model.

One may define the Euler-Heisenberg Lagrangian, $\mathbb{L}_\text{EH}[A,A_5]$ with axial gauge, from the one-loop effective action,
\begin{equation}\label{eq:eff_action_def}
	\Gamma[A,A_5]=-i\textrm{Tr}\ln[i\gamma^{\mu}(D_{\mu}+i\gamma_{5}A_{5\mu})-m]\,,
\end{equation}
in homogeneous fields for system spacetime volume, $\mathcal{TL}^3$ (where $\mathcal{T} $ and $\mathcal{L} $ are the system duration and spatial length, respectively) as $\Gamma[A,A_5]=\mathcal{TL}^3 \mathbb{L}_\text{EH}[A,A_5]-\Gamma_{\text{Maxwell}}$,   including the Maxwell Lagrangian (action). 
To study the one-loop QED effective action with an axial gauge we make use of the propertime method~\cite{Fock:1937dy,*Nambu:1950rs,*Feynman:1950ir, *PhysRev.82.664}, and evaluate using a spectral decomposition of the fermionic determinant~\cite{fradkin1991quantum,schwartz_2013}, as well as the worldline instanton method~\cite{AFFLECK1982509,*PhysRevD.72.105004}.  The former method relies on expressing the effective action as a sum over its eigenvalues, and provides exact solutions for, e.g., the case of a homogeneous~\cite{PhysRevD.58.105022} or Sauter background~\cite{fradkin1991quantum}.  Here we exploit two special cases in which exact eigenvalues of the Dirac operator can be found by virtue of the Ritus basis~\cite{RITUS1972555,*Ritus:1978cj}: a magnetic field with chiral chemical potential and an electric field with spatial axial gauge.  Note that a novel feature here is that we extend the Ritus basis to an equivalent formulation for electric fields. What is more, is we determine that an enhancement of Schwinger pair production can be found with a spatial axial gauge.  To support this finding we study it semi-classically as well using the worldline instanton method~\cite{AFFLECK1982509,*PhysRevD.72.105004}; what is novel here as well is that we adopt the method for the phase-space worldline formalism~\cite{Migdal:1986pz,*doi:10.1142/7305,*Fosco:2003rr}.  The worldline instanton method has made possible the study of pair production in otherwise complicated backgrounds, not limited to dynamically assisting fields~\cite{PhysRevLett.101.130404}, under finite temperature~\cite{PhysRevD.95.056006,*PhysRevD.96.076002,*PhysRevD.98.056022},  with non-Abelian backgrounds~\cite{PhysRevD.103.036004},  and in the massive Schwinger model~\cite{schwinger_model_pair_production}. Euler-Heisenberg Lagrangians have been studied perturbatively in electromagnetic and axial gauge couplings~\cite{Yamashita:2017scc,*Ghasemkhani:2021kzf}, and related worldline approaches have been put forth that include an axial gauge field~\cite{MONDRAGON1996212,*DHOKER1996297,*DHOKER1996272}.

We first analyze the linear Dirac operator's eigenvalues and how they lead to a perturbative in the $\mathcal{O}(A_5)$ definition of the effective action in Sec.~\ref{sec:eigenvalues}. Next we take a deeper look at the Schwinger pair production enhancement  for spatial axial gauges through the worldline instanton method in Sec.~\ref{sec:worldline}.  Last, we treat the exactly solvable theory for the massless case, whether massive in the massless limit or purely massless in Sec.~\ref{sec:mass}.

A note on our conventions: We use for our Minkowski metric $\eta^{\mu\nu}=\textrm{diag}(1,-1,-1,-1)$, and for the completely antisymmetric tensor, $\varepsilon^{\mu\nu\alpha\beta}$, we take $\varepsilon^{0123}=1$.  Our covariant derivative reads $D_\mu=\partial_\mu + iqA_\mu$. Also we take $\gamma_{5}=i\gamma^{0}\gamma^{1}\gamma^{2}\gamma^{3}$ and use Weyl matrices:
\begin{equation}
	\gamma_{0}=\Bigr(\begin{array}{cc}& I_{2}\\I_{2}\end{array}\Bigl),\;
	\gamma^{i}=\Bigr(\begin{array}{cc}& \sigma^{i}\\-\sigma^{i}\end{array}\Bigl),\;
	\gamma_{5}=\Bigr(\begin{array}{cc}-I_{2}\\& I_{2}\end{array}\Bigl)\,.
\end{equation}
Let us define the spin tensor as $\sigma_{\mu\nu}=(i/2)[\gamma_{\mu},\gamma_{\nu}]$. Finally where appropriate, we make use of a matrix form in Lorentz indices, i.e., $F^\mu_{\;\nu}=F$ and $x^\mu=x$, with contractions assumed in place of matrix multiplication.

\section{Spectral Decomposition}
\label{sec:eigenvalues}

We begin our evaluation of the massive spinor one-loop effective action (alternatively the Euler-Heisenberg Lagrangian~\cite{Heisenberg:1935qt} with axial gauge) by remarking that the argument of the associated functional determinant from the fermion partition function has a formal eigendecomposition as follows:
\begin{equation}\label{eq:operator}
   \hat{\slashed{\Pi}}_5|\psi_N\rangle\equiv i\gamma^{\mu}(D_{\mu}+i\gamma_{5}A_{5\mu})|\psi_N\rangle=\lambda_N|\psi_N\rangle\,,
\end{equation}
where $\lambda_{N}$ spans the entire eigenspectrum with quantum number, $N$.  The eigendecomposition allows for the transformation of the effective action, Eq.~\eqref{eq:eff_action_def}, as a sum over its eigenvalues as
\begin{align}\label{eq:eff_action_single_1}
\Gamma[A,A_{5}]&=-i\textrm{tr}\int d^{4}x\langle x| \ln[\hat{\slashed{\Pi}}_5-m]|x\rangle\nonumber \\
 & =-i \sum_N\ln[\lambda_N-m]\,.
\end{align}
We have inserted in a complete set of eigenstates, $\sum_N|\psi_N\rangle\langle\psi_N|=1$,
and assumed a normalization of $\textrm{tr}\int d^{4}x\langle x|\psi_N\rangle\langle\psi_N|x\rangle=1$.
The mass $m\rightarrow m-i\epsilon$ can provide convergence for a
Schwinger propertime integral; in this sense the log of operator is
understood. By making use of the fact that $\gamma_5$ anti-commutes with the operator given in Eq.~\eqref{eq:operator}, one may also equally well write 
\begin{align}
\Gamma[A,A_{5}] & =-i\textrm{Tr}\ln[\gamma_{5}\hat{\slashed{\Pi}}_5\gamma_{5}-m]\nonumber \\
 & =-i\sum_N\ln[-\lambda_N-m]\,.
\end{align}
Then averaging over both the above and Eq.~\eqref{eq:eff_action_single_1}, and through the application of Schwinger propertime~\cite{PhysRev.82.664}, we arrive at an alternative expression of the effective action as the sum over quadratic eigenvalues as
\begin{equation}\label{eq:quad_eigenvalue_sum}
\Gamma[A,A_{5}] =\frac{i}{2}\int_{0}^{\infty}\frac{dT}{T}\sum_Ne^{-i(-\lambda_N^{2}+m^{2})T}\,.
\end{equation}
Then determination of the full effective action has been reduced to the evaluation of eigenvalues, $\lambda_N$, for $\langle x|\psi_N\rangle=\psi_N$ where
\begin{equation}
\hat{\slashed{\Pi}}_5 \psi_N=\lambda_N \psi_N \,.\label{eq:eigenvalue_eq}
\end{equation}

An intuitive strategy for the evaluation of Eq.~\eqref{eq:eigenvalue_eq} in homogeneous electromagnetic fields is to seek a separable solution wherein modes for the magnetic (electric) field are described by their Landau levels (electric equivalent imaginary Landau levels). However, due to the presence of the axial gauge this separation is not easily implemented; in principal $A_5^\mu$ cannot be removed by a field redefinition. (This is, however, possible in the purely massless case; see~\ref{sec:mass}.) Even so, there are two particular background electromagnetic and axial gauge configurations where separable solutions are readily found, one for each set of Landau levels.  These cases are the following:
\begin{enumerate}
\item A magnetic field with a chiral chemical potential:
\begin{equation}\label{eq:mag_field}
A_{\mu}=\frac{B}{2}(\delta_{\mu}^{1}x^{2}-\delta_{\mu}^{2}x^{1})\,,\quad A_{5}^{\mu}=g^{\mu0}\mu_{5}\,.
\end{equation}
\item An electric field with spatial axial gauge:
\begin{equation}\label{eq:elec_field}
A_{\mu}=\frac{E}{2}(\delta_{\mu}^{0}x^{3}-\delta_{\mu}^{3}x^{0})\,,\quad A_{5}^{\mu}=g^{\mu1}\omega_{5}\,.
\end{equation}
\end{enumerate}
In both cases a Fock-Schwinger gauge has been used, $A_{\mu}=-\frac{1}{2}F_{\mu\nu}x^{\nu}$. Let us, however, stress that the eigenvalues of Eq.~\eqref{eq:eigenvalue_eq} are gauge-invariant since the electromagnetic part of the spectrum is determined by $[D_\mu,D_\nu]=iqF_{\mu\nu}$; one may equally well use another electromagnetic gauge, which is a transform of the above. For case 2, an essential identification of the spacelike axial gauge is its connection to vorticity~\cite{PhysRevD.103.125003}.  This identification proceeds along the lines of vorticity$\rightarrow$torsion in spacetime~\cite{PhysRevB.90.134510,*Khaidukov2018}$\rightarrow$axial gauge field~\cite{SHAPIRO2002113}; we refer the reader to~\cite{PhysRevD.103.125003} for a derivation along with supplementary details. Let us denote for the strength of vorticity in a local rest frame as $\omega$, where $\bm \omega=\bm\nabla\times \bm v/2$ with $\bm v$ being a rotational velocity of the system and $\omega=|\bm \omega|$. It is, however, more convenient for us to absorb a factor of one-half, i.e., $\omega=2\omega_5$ as in Eq.~\eqref{eq:elec_field} in this work, and therefore we limit our discussions to $\omega_5$.
Let us first proceed with the evaluation of the eigenvalues for the magnetic field case next.

\subsection{Magnetic field with chiral chemical potential}
\label{sec:magnetic_eigenvalue}

The selection of a magnetic field with chiral chemical potential, Eq.~\eqref{eq:mag_field}, we will show benefits from an exactly solvable setup by virtue of the Ritus basis~\cite{RITUS1972555,*Ritus:1978cj}.  Note our conventions follow those used in~\cite{Hattori:2020htm,Hattori_review}.  To begin our discussion let us first introduce magnetic field spin projection operators:
\begin{equation}
\mathcal{P}_{\pm}\coloneqq\frac{1}{2}\bigl(1\pm\sigma^{12}\textrm{sgn}(qB)\bigr)\,.
\end{equation}
The projection operators satisfy idempotency, completeness, and orthogonality,
i.e., $\mathcal{P}_{\pm}\mathcal{P}_{\pm}=\mathcal{P}_{\pm}$, $\mathcal{P}_{+}+\mathcal{P}_{-}=1$,
and $\mathcal{P_{\pm}\mathcal{P}_{\mp}}=0$. We also have that $\gamma^{1,2}\mathcal{P}_{\pm}=\mathcal{P}_{\mp}\gamma^{1,2}$. Furthermore,  the spin projection operators commute with terms in the eigenvalue equation containing the chiral chemical potential. The magnetic field spin projection operators act to diagonalize the spin factor associated with the magnetic field as $qB\sigma_{12}\mathcal{P}_{\pm}=\pm|qB|\mathcal{P}_{\pm}$.

Before introducing the Ritus basis it is convenient to highlight the harmonic
oscillator parallels of the Dirac equation in a magnetic field.  Notably
with the eigenvalue (Dirac) equation, Eq.~\eqref{eq:eigenvalue_eq}, in a magnetic field with chiral chemical potential one has
\begin{equation} \label{eq:Dirac_BMu}
 i\Bigl[\gamma^{0}(\partial_{0}+\gamma_{5}\mu_{5})-\gamma^{1}D^{1}-\gamma^{2}D^{2}+\gamma^{3}\partial_{3}\Bigr]\psi_N=\lambda_N\psi_N\,;
\end{equation}
notice that we may introduce the following creation and annihilation operators:
\begin{align}
\hat{a}&=\frac{1}{\sqrt{2|qB|}}[iD^{1}-\textrm{sgn}(qB)D^{2}]\,,\\
\hat{a}^{\dagger}&=\frac{1}{\sqrt{2|qB|}}[iD^{1}+\textrm{sgn}(qB)D^{2}]\,,
\end{align}
that satisfy $[\hat{a},\hat{a}^{\dagger}]=1$, $[\hat{a},\hat{a}]=[\hat{a}^{\dagger},\hat{a}^{\dagger}]=0$.
A quantum harmonic oscillator solution is then available, $\hat{a}^{\dagger}\hat{a}\langle x|n\rangle=n\langle x|n\rangle = n\phi_{n}$.  
One may then write for the magnetic Hamiltonian, $-(D^{1})^{2}-(D^{2})^{2}=(2\hat{a}^{\dagger}\hat{a}+1)|qB|$.
Then it immediate follows that the eigenvalues are
$(2n+1)|qB|$ for $n=0,1,2...$, which will lead to the Landau levels. 
We can also determine that for
the normalized eigenbasis one must have that $\hat{a}\phi_{n}=\sqrt{n}\phi_{n-1}$
and $\hat{a}^{\dagger}\phi_{n}=\sqrt{n+1}\phi_{n+1}$.  One may then rewrite for the argument in the eigenvalue equation
\begin{equation}
-i\gamma^{1}D^{1}-i\gamma^{2}D^{2}=-\sqrt{2|qB|}\gamma^{1}(\hat{a}\mathcal{P}_{+}+\hat{a}^{\dagger}\mathcal{P}_{-})\,.
\end{equation}

Let us now introduce the Ritus basis to evaluate the magnetic degrees of freedom in the eigenvalue equation. The basis is
\begin{equation}\label{eq:Ritus}
\mathcal{R}_{n}(x_{1,2})=\phi_{n}(x_{1,2})\mathcal{P}_{+}+\phi_{n-1}(x_{1,2})\mathcal{P}_{-}\,,
\end{equation}
which combines the spin degenerate Landau levels.  Note that a lower truncation of $\phi_{-1}\equiv0$ is assumed. The Ritus basis acts to diagonalize the quadratic operator as $[-(D^{1})^{2}-(D^{2})^{2}-\sigma^{12}qB]\mathcal{R}_{n}(x_{1,2}) =2n|qB|\mathcal{R}_{n}(x_{1,2})$. Let us also point out that the magnetic eigenvectors also have a perpendicular canonical momentum dependence acting to shift the center of the wavepacket; we will introduce the quantum label where necessary. The basis allows us to evaluate the eigenvalue equation, Eq.~\eqref{eq:Dirac_BMu}, for arbitrary spinor solution, $z_s$, as
\begin{equation}
\psi_{N}(x)=e^{-ip^{0}x^{0}+ip^{3}x^{3}}\mathcal{R}_{n}(x_{1,2})z_s\,.
\end{equation}
Inserting the above into Eq.~(\ref{eq:Dirac_BMu}), the following relationship can be found:
\begin{align}\label{eq:magnetic_characteristic}
\mathcal{R}_{n}\Bigl[\gamma^{0}(p^{0}-\gamma_{5}\mu_{5})-\sqrt{2|qB|n}\gamma^{1}-\gamma^{3}p^{3}-\lambda\Bigr]z_s=0\,.
\end{align}
A non-trivial solution to the above can be had if the expression in brackets, along with the spinor acting on it, vanishes.  We can see that solutions are simply a modified Dirac equation in a chiral chemical potential; the chiral chemical potential lifts the degeneracy of eigenvalues and now four independent eigenvalues can be found reading
\begin{equation}\label{eq:magnetic_eigenvalues}
\lambda_N=\pm\sqrt{(p^{0})^{2}-p_n^{\parallel 2}-\mu_{5}^{2}+2s|\mu_{5}|p_n^\parallel}\,,
\end{equation}
where $p_n^\parallel = \sqrt{2|qB|n+(p^{3})^{2}}$ 
and $s=\pm1$ is an eigenvalue of the matrix ${\mathcal S} = -  \gamma_{5} \gamma^{0} ( \sqrt{2|qB|n}\gamma^{1} + \gamma^{3}p^{3})/p_n^\parallel  $. 
In the case of the lowest Landau level (LLL), those expressions reduce to 
$\lambda_N = \pm\sqrt{ (p^0)^2   - (|p^3| - \mu_5 s)^2 }$ and ${\mathcal S}= - {\rm sgn}(p^3)  i \gamma^1 \gamma^2$ that is the helicity operator. 
Since we will be summing over the squared operator there will be an overall degeneracy associated with squaring each eigenvalue above. Therefore we can conclude based on the above that the set of good quantum numbers includes
\begin{equation}
    N=\{p^0,p^3,n,s\}\,.
\end{equation}

With the above eigenvalues one may determine the effective action, Eq.~\eqref{eq:quad_eigenvalue_sum}.  
It should be noticed that, in addition to $ n$, 
there should be another quantum number so that the number of degrees of freedom remains the same as in the case without the magnetic field. 
This quantum number does not appear in the eigenvalue (\ref{eq:magnetic_eigenvalues}), indicating energy degeneracy. 
The density of degenerate states can be counted conveniently by using the Landau gauge $A_{\mu}= \delta_{\mu}^{1}Bx^{2} $ which can be thought of as the spatial rotation of the Fock-Schwinger gauge (\ref{eq:mag_field}) by the use of the gauge. 
Then, one finds that one component of the canonical momentum 
$p^c = p^1 $ is a good quantum number. 
Therefore, the summation over the quantum numbers is implemented as 
\begin{equation}
\sum_N \rightarrow \frac{\mathcal{T}\mathcal{L}^2\int dp^{0,3,\textrm{c}}}{(2\pi)^{3}} \sum_{n,s}\,,
\end{equation}
The integration over the modes is unbounded in homogeneous magnetic fields, and one can determine the physical cutoff by considering a closed box of Landau modes~\cite{schwartz_2013}. This leads to $\int dp^\textrm{c} = |qB|\mathcal{L}$. 
The physics behind this is that the degeneracy stems from the distribution of cyclotron orbits in the transverse plane and the conserved canonical momentum serves as the center coordinate $ \xi$ of the orbits via the relation $\xi = p^c/|qB|$, leading to a constraint $0 \leq p^c \leq  |qB|\mathcal{L} $. Last, there is an overall factor of 2 to account for the degeneracy of both the positive and the  negative components of the eigenvalues in $\lambda_N$ after squaring both.

Let us next sum over the Landau levels. A virtue of the Ritus basis is that it combines the spin degenerate Landau levels into one value of $n$, and the lowest Landau level is accounted for under $\phi_{-1}=0$. We can see this in the construction of the basis in Eq.~\eqref{eq:Ritus}, one eigenvector for $n$ and another $n-1$, and both are independent of one another guaranteed by the spin projectors. The key point here is that we must have a factor of one-half for $n=0$ in comparison to the higher Landau levels, $n>0$, since only the $\mathcal{P}_+$ factor is present. Gathering everything together, including the volume and momentum integrals, we find:

\begin{align}\label{eq:eff_B_sum_eigen}
&\Gamma(B,\mu_{5})=\frac{i}{4\pi}\mathcal{TL}^{3}|qB|\int_{0}^{\infty}\frac{dT}{T}\int\frac{dp^{0,3}}{(2\pi)^{2}}\,e^{-im^2T} \sum_{s=\pm1} \nonumber \\
&\times 2\Bigl\{ \sum_{n=1}^{\infty}e^{i(\lambda_{p^0,p^3,n,s})^{2}T}+\frac{1}{2}e^{i(\lambda_{p^0,p^3,0,s})^{2}T}\Bigr\}\,.
\end{align}   
Let us furthermore evaluate the $p^0$ integral and take the sum over $s$ to find
\begin{align}
&\Gamma(B,\mu_{5})=\frac{i\mathcal{TL}^{3}|qB|}{(2\pi)^{3}}\int_{0}^{\infty}\frac{dT}{T}\sqrt{\frac{\pi i}{T}}\int dp^{3}e^{-i(m^{2}+\mu_{5}^{2})T}\nonumber \\
&\times\Bigl\{2\sum_{n=1}^{\infty}e^{-ip_{n}^{\parallel 2}T}\cos(2\mu_{5}p_{n}^{\parallel}T)+e^{-ip_{3}^{2}T}\cos(2\mu_{5}|p_3|T)\Bigr\}\,.
\end{align}
To proceed let us take a power series expansion of the cosine terms expressing the arguments as propertime partials, then integrating over $p^3$, and last summing over the Landau levels as $2\sum_{n=1}^{\infty}e^{-2inx}+1=-i\cot x$ to find
\begin{align}\label{eq:eff_B_diff}
&\Gamma(B,\mu_{5})=\frac{\mathcal{TL}^{3}|qB|}{8\pi^{2}}\int_{0}^{\infty}\frac{dT}{T}\sqrt{\frac{1}{T}}e^{-i(m^{2}+\mu_{5}^{2})T}\nonumber \\
&\times\sum_{k=0}^{\infty}\frac{(-1)}{(2k)!}^{k}(2\mu_{5}T)^{2k}i^{k}\partial_{T}^{k}\sqrt{\frac{1}{T}}\cot(|qB|T)\,.
\end{align}
To clarify $\partial_T^0\equiv 1$. In either of the above representations, whether in integral or summation form,  the effective action is still exact.  However, in the above expression one may perform a convenient perturbation in $\mu_5$ analysis.  Let us therefore evaluate Eq.~\eqref{eq:eff_B_diff} to $\mathcal{O}(\mu_5^4)$ to find
\begin{align}\label{eq:eff_B_pert}
&\Gamma(B,\mu_{5})\approx \frac{\mathcal{TL}^{3}|qB|}{8\pi^{2}}\int_{0}^{\infty}dT \Bigl\{\frac{c_B}{T^2}+2i\mu_{5}^{2}|qB|(1+c_{B}^{2})\nonumber \\
&+\mu_{5}^{4}\frac{4}{3}|qB|(1+c_{B}^{2})T[1-|qB|c_{B}T]\Bigr\}e^{-im^{2}T}\,,
\end{align}
where $c_B\equiv \cot(|qB|T)$.

One can see in Eq.~\eqref{eq:eff_B_diff}, and more readily in Eq.~\eqref{eq:eff_B_pert} that poles only reside on the real propertime axis. Therefore we can conclude that there should not be an imaginary contribution to the effective action with the presence of $\mu_5$ (and a magnetic field) since all poles can be associated with a power of $c_B$, and hence there would not be Schwinger pair production as in the sole magnetic field case.

Let us evaluate the real part of Eq.~\eqref{eq:eff_B_pert}. First let us explore the case of a weak magnetic field, to $\mathcal{O}(|qB|^4)$. There is a UV divergence as there is in the case without the chiral chemical potential. Let us treat this with a physical cutoff at a QED scale of $\Lambda^{-2}$ for small Schwinger propertimes.  To find the adequate effective representation we first deform the contour to the imaginary propertime axis and perform a change of variables as $T\rightarrow -iT$, and then we introduce the cutoff.  We find for small magnetic fields,
\begin{align}
\Gamma(B,\mu_{5})&=\frac{\mathcal{TL}^{3}}{8\pi^{2}}\int_{\Lambda^{-2}}^{\infty}\frac{dT}{e^{m^{2}T}}\Bigl\{-\frac{1}{T^{3}}-\frac{|qB|^{2}}{3T}+\frac{|qB|^{4}T}{45}\notag \\
&+2\mu_{5}^{2}\Bigl(-\frac{1}{T^{2}}+\frac{|qB|^{2}}{3}-\frac{|qB|^{4}T^{2}}{15}\Bigr)\notag\\
&+\frac{4}{9}\mu_{5}^{4}\Bigl(-|qB|^{2}T+\frac{2}{5}|qB|^{4}T^{3}\Bigr)\Bigr\}\,.
\end{align}
The term proportional to $T^{-3}$ is independent of the fields and axial gauge and should be removed by a suitable counterterm in the effective action, i.e., $\Gamma_{\text{reg.}}[A,A_5]=\Gamma[A,A_5]-\Gamma[0,0]$. Let us emphasize that we treat the axial gauge field as a background field. 
Since we have treated terms to both $\mathcal{O}(\mu_5^4)$ and $\mathcal{O}(|qB|^4)$ we may safely ignore products of $\mu_5$ and $|qB|$ whose combined order is greater than four, e.g., $\mathcal{O}(\mu_5^4|qB|^2)$ such as in the last line; we will find such terms are in any case not divergent. For later usage let us write out each integral from $T^3$ to $T^{-3}$:
\begin{align}
&\int_{\Lambda^{-2}}^{\infty}T^kdTe^{-m^{2}T}\approx \frac{k!}{m^{2(k+1)}}, \quad k=0,1,2,3\,,\label{eq:int_1}\\
&\int_{\Lambda^{-2}}^{\infty}\frac{dT}{T}e^{-m^{2}T}\approx-\ln\bigl(\frac{m^{2}}{\Lambda^{2}}\bigr)\,,\\
&\int_{\Lambda^{-2}}^{\infty}\frac{dT}{T^{2}}e^{-m^{2}T}\approx\Lambda^{2}+m^{2}\ln\bigl(\frac{m^{2}}{\Lambda^{2}}\bigr)\,,\\
&\int_{\Lambda^{-2}}^{\infty}\frac{dT}{T^{3}}e^{-m^{2}T}\approx\frac{\Lambda^{4}}{2e^{\frac{m^2}{\Lambda^2}}}-\frac{m^{2}}{2}\bigl[\Lambda^{2}+m^{2}\ln\bigl(\frac{m^{2}}{\Lambda^{2}}\bigr)\bigr]\,,\label{eq:int_4}
\end{align}
where we have kept only leading order terms in $\Lambda^2$. Then one can find that
\begin{equation}\label{eq:eff_free}
\Gamma(0,0)=\frac{\mathcal{TL}^{3}}{8\pi^{2}}\Bigl\{-\frac{1}{2}\Lambda^{4}+m^{2}\Bigl[\Lambda^{2}+\frac{m^{2}}{2}\ln\bigl(\frac{m^{2}}{\Lambda^{2}}\bigr)\Bigr]\Bigr\}\,.
\end{equation}
Moreover we have for weak magnetic fields and chiral chemical potential that
\begin{align}\label{eq:eff_real_magnetic}
\Gamma&_\text{reg.}(B,\mu_{5})=\frac{\mathcal{TL}^{3}}{8\pi^{2}}\Bigl\{\frac{|qB|^{2}}{3}\ln\bigl(\frac{m^{2}}{\Lambda^{2}}\bigr)+\frac{1}{45}\frac{|qB|^{4}}{m^{4}}\notag\\
&+2\mu_{5}^{2}\Bigl(-\Lambda^{2}-m^{2}\ln\bigl(\frac{m^{2}}{\Lambda^{2}}\bigr)+\frac{|qB|^{2}}{3m^{2}}\Bigr)\Bigr\}\,.
\end{align}
One can see that Eq.~\eqref{eq:eff_real_magnetic} is dominant in $\Lambda^2$. Hence we can tell with the presence of a sole chiral chemical potential the energy of the system is decreased, indicating stability. Also, notice that an augmentation of the quadratic $|qB|^2$ term is present; this suggests that the axial gauge field also augments the Maxwell electromagnetic Lagrangian, affecting photon-propagation~\cite{Adler:1971wn}.

To extract physically relevant quantities one must define a renormalization scheme. We follow an approach as employed for the usual Euler-Heisenberg Lagrangian~\cite{dittrich1985effective}, however, with the inclusion of an axial gauge field. In addition to the usual divergence associated with two insertions of the magnetic field to the fermion loop, we also have a divergence with $\mu_5$ as can be seen in~\eqref{eq:eff_real_magnetic}. Nevertheless, one can find renormalization of $q$ and $B$ that incorporates $\mu_5$. Let us be more specific. The $q$ and $B$ that have been used so far are bare (and un-physical) quantities. Let us denote those bare quantities as $q_0$ and $B_0$ instead, and the renormalized and physical ones as $q$ and $B$, respectively, such that $B=B_0/\sqrt{Z_3}$ and $q=q_0\sqrt{Z_3}$, and $qB=q_0B_0$. Then one can find that the Euler-Heisenberg Lagrangian under the $\overline{{\rm MS}}$ scheme reads 
\begin{align}
\label{eq:L-renom}
    \mathbb{L}_\text{EH}&=-\frac{B^2}{2}+ \frac{1}{8\pi^{2}}\int_{0}^{\infty}\frac{dT}{e^{im^{2}T}} \Bigl\{\frac{|qB|c_B}{T^2}+\frac{|qB|^2}{3T}-\frac{1}{T^3}\notag\\
    &+2i\mu_{5}^{2}\Bigl[|qB|^2(1+c_{B}^2)-\frac{1}{T^2}\Bigr]\Bigr\}\,,
\end{align}
where the renormalization factor is found to be 
\begin{equation}
Z_3=1+\frac{q_0^2}{12\pi^2}\int^{\infty}_{\Lambda^{-2}}\frac{dT}{T}e^{-im^2T}\Big(1-\frac{6i\mu_5^2}{(q_0B_0)^2T}\Big)\,.
\end{equation}
Alternatively, one can think of the above renormalization scheme as the subtraction of the vacuum energy with 
$\bar \Gamma_{\text{reg.}}[A,A_5]=\Gamma[A,A_5]-\Gamma[0,A_5]$. 
Unlike in~\eqref{eq:eff_real_magnetic}, 
the regularized effective Lagrangian no longer has a divergence associated with $\mu_5$ 
as well as those in \eqref{eq:eff_real_magnetic}. 
Then, we arrive at the same form of the renormalized effective Lagrangian (\ref{eq:L-renom}) by applying the $\overline{{\rm MS}}$ scheme to the divergence associated with the two insertions of the magnetic field only. Having explored the case of a magnetic field with chiral chemical potential let us turn our attention to the case of an electric field with spatial axial gauge.

\subsection{Electric field and spatial axial gauge}
\label{sec:electric_eigenvalue}

Let us now address the case of an electric field with spatial axial gauge as given in Eq.~\eqref{eq:elec_field}. With strong similarity to the above case with a magnetic field, we derive a Ritus-like basis for the electric field,  whose eigenvalues, or electric-like Landau levels, are now imaginary. Much of the discussion here parallels the one in Sec.~\ref{sec:magnetic_eigenvalue}; however, with important subtleties. The final form of our effective action can be seen in Eq.~\eqref{eq:ea2_final}. First, let us introduce electric field spin projection operators:
\begin{equation}
\mathrm{P}_{\pm}\coloneqq\frac{1}{2}\bigl(1\pm i\sigma^{03}\textrm{sgn}(qE)\bigr)\,.
\end{equation}
They satisfy as before $\mathrm{P}_{\pm}\mathrm{P}_{\pm}=\mathrm{P}_{\pm}$, $\mathrm{P}_{+}+\mathrm{P}_{-}=1$,
and $\mathrm{P}_{\pm}\mathrm{P}_{\mp}=0$. We also have that $\gamma^{0,3}\mathrm{P}_{\pm}=\mathrm{P}_{\mp}\gamma^{0,3}$. Furthermore, we have that $[\textrm{P}_{\pm},\gamma^{1,2}]=0$. The operators have the virtue of projecting eigenvalues of the spin
matrices as $-qE\sigma^{03}\mathrm{P}_\pm =\pm i|qE|\mathrm{P}_{\pm}$.

The eigenvalue equation in the electric field with spacelike axial gauge becomes
\begin{equation}\label{eq:Dirac_electric}
i\Bigl[\gamma^{0}D^{0}+\gamma^{1}(\partial_{1}+\gamma_{5}\omega_{5})+\gamma^{2}\partial_{2}-\gamma^{3}D^{3}\Bigr]\psi_N=\lambda_N\psi_N\,.
\end{equation}
As before we will find that the above differential equation can be solved through the introduction of creation and annihilation operators:
\begin{align}
\bar{\textrm{a}}&=\frac{1}{\sqrt{-2i|qE|}}[D^{0}-\textrm{sgn}(qE)D^{3}]\,,\\
\mathrm{a}&=\frac{1}{\sqrt{-2i|qE|}}[D^{0}+\textrm{sgn}(qE)D^{3}]\,,
\end{align}
which satisfy $[\mathrm{a},\bar{\mathrm{a}}]=1$ and $[\mathrm{a},\mathrm{a}]=[\bar{\mathrm{a}},\bar{\mathrm{a}}]=0$. Note that for the Ritus basis for the electric field $\bar{\textrm{a}}\neq\textrm{a}^\dagger$. We now have that $(D^{0})^{2}-(D^{3})^{2} =-(2\bar{\mathrm{a}}\mathrm{a}+1)i|qE|$, and to determine the eigenspectrum let us note that $\bar{\mathrm{a}}\mathrm{a}|\textrm{n}\rangle  =\textrm{n}|\textrm{n}\rangle$,  $\bar{\mathrm{a}}\mathrm{a}\bar{\mathrm{a}}|\textrm{n}\rangle =(\textrm{n}+1)\bar{\mathrm{a}}|\textrm{n}\rangle$, and $\bar{\mathrm{a}}\mathrm{a}\mathrm{a}|\textrm{n}\rangle =(\textrm{n}-1)\mathrm{a}|\textrm{n}\rangle$. Therefore we can find the eigenspectrum of $(D^{0})^{2}-(D^{3})^{2}$
as the electric field equivalent of the Landau levels: $-(2\textrm{n}+1)i|qE|\,\forall \textrm{n}\in\mathbb{Z}^{0+}$, where we notice the eigenvalues are now imaginary.  Also as an immediate consequence for normalization we have that $\mathrm{a}\phi_{\textrm{n}}=\sqrt{\textrm{n}}\phi_{\textrm{n}-1}$ and $\bar{\mathrm{a}}\phi_{\textrm{n}}=\sqrt{\textrm{n}+1}\phi_{\textrm{n}+1}$. Finally we can express the linear Dirac operator in the following form: 
\begin{equation}
i\gamma^{0}D^{0}-i\gamma^{3}D^{3}=i\sqrt{-2i|qE|}\gamma^{0}[\bar{\mathrm{a}}\mathrm{P}_{-}+\mathrm{a}\mathrm{P}_{+}]\,.
\end{equation}

We may now define the Ritus-like basis for the electric fields as 
\begin{equation}
\mathrm{R}_{\textrm{n}}(x_{0,3})=\phi_{\textrm{n}}(x_{0,3})\mathrm{P}_{+}+\phi_{\textrm{n}-1}(x_{0,3})\mathrm{P}_{-}\,,
\end{equation}
where we take that $\phi_{-1}\equiv0$.  As before this basis diagonalizes the quadratic Dirac operator with electric fields as $[-(D_{0})^{2}+(D_{3})^{2}+qE\sigma^{03}]\mathrm{R}_{\textrm{n}}(x_{0,3})=2i|qE|\textrm{n}\mathrm{R}_{\textrm{n}}(x_{0,3})$. One can then determine a solution to the Dirac equation as
\begin{equation}
\psi_{p^1,p^2,\textrm{n},s}(x)=e^{ip^{1}x^{1}+ip^{2}x^{2}}\mathrm{R}_{\textrm{n}}(x_{0,3})z_{s}\,,
\end{equation}
for arbitrary spinor $z_s$.  From the eigenvalue equation, Eq. (\ref{eq:Dirac_electric}), using the above
we can now find that the following expression must be met
\begin{equation}
\mathrm{R}_{\textrm{n}}\Bigl[i\sqrt{-2i|qE|m}\gamma^{0}-\gamma^{1}(p^{1}+\gamma_{5}\omega_{5})-\gamma^{2}p^{2}-\lambda_N\Bigr]z_{s}=0\,.
\end{equation}
A non-trivial solution can be found when the expression in the brackets acting on the spinor vanishes. We find the four independent eigenvalues are
\begin{equation}\label{eq:electric_eigenvalues}
\lambda_{N}=\pm\sqrt{p_\textrm{n}^{\parallel 2}-p_{1}^{2}+\omega_{5}^{2}+2s|\omega_{5}|p_\textrm{n}^\parallel}\,.
\end{equation}
where we write $p_\textrm{n}^\parallel=\sqrt{2i|qE|\textrm{n}-p_{2}^{2}}$, and once again $s=\pm 1$.

Using Eq.~\eqref{eq:quad_eigenvalue_sum}, one may now express the effective action as a sum over the above eigenvalues in the same way as before as
\begin{align}\label{eq:ea2_final}
&\Gamma(E,\omega_{5})=\frac{i}{4\pi}\mathcal{T}\mathcal{L}^{3}|qE|\int_{0}^{\infty}\frac{dT}{T}\int\frac{dp^{1,2}}{(2\pi)^{2}}\sum_{s=\pm1}e^{-im^{2}T}\nonumber \\
 &\times2\Bigl\{\sum_{\textrm{n}=1}^{\infty}e^{i(\lambda_{p^1,p^2,\textrm{n},s}^2)T}+\frac{1}{2}e^{i(\lambda_{p^1,p^2,0,s}^2)T}\Bigr\}\,.
\end{align}
The above effective action is exact albeit in integral form. Let us proceed as before with its perturbative in $\omega_5$ evaluation. To do so we integrate out $p^1$ and sum over $s$ to find
\begin{align}
&\Gamma(E,\omega_{5})=\frac{i\mathcal{TL}^{3}|qE|}{8\pi^{3}}\int_{0}^{\infty}\frac{dT}{T}\sqrt{\frac{\pi}{iT}}\int dp^{2}\,e^{-i(m^{2}-\omega_5^2)T}\nonumber \\
 &\times\Bigl\{2\sum_{\textrm{n}=1}^{\infty}e^{ip_\textrm{n}^{\parallel 2}T}\cos(2\omega_{5}p_\textrm{n}^\parallel T)+e^{-ip_{2}^{2}T}\cos(2i\omega_{5}|p_{2}|T)\Bigr\}\,.
\end{align}
Expanding out the cosine terms in a power series whose argument we express as a propertime partial derivative, and summing over the electric Landau levels as $2\sum_{m=1}^{\infty}e^{-2mx}+1=\coth(x)$, we find the effective action is expressible as
\begin{align}\label{eq:eff_spatial_diff}
&\Gamma(E,\omega_{5})=\frac{\mathcal{TL}^{3}|qE|}{8\pi^{2}}\int_{0}^{\infty}\frac{dT}{T^{3/2}} \,e^{-i(m^{2}-\omega_{5}^{2})T}\nonumber \\
&\times\sum_{k=0}^{\infty}\frac{(-1)^{k}}{(2k)!}(2\omega_{5}T)^{2k}(-i)^{k}\partial_{T}^{k}\sqrt{\frac{1}{T}}\coth(|qE|T)\,.
\end{align}

Let us proceed with the evaluation of the above for its imaginary part, which we will show one may express in summation form. 
The integrand in Eq.~(\ref{eq:eff_spatial_diff}) has $n$ poles at $T=-in\pi/|qE|\,\forall n\in\mathbb{Z}^+$ from the hyperbolic cotangent function, and hence one may evaluate the imaginary part of the effective action by deforming the Schwinger propertime contour to the lower half of the complex $T$ plane and applying the residue theorem. Summing the contributions of $n$ poles we may write
\begin{equation}\label{eq:Im_sum_poles}
    \mathrm{Im}\Gamma(E,\omega_{5})=-\pi\frac{\mathcal{TL}^{3}|qE|}{8\pi^{2}}\sum_{n=1}^\infty\sum_{k=0}^\infty \mathrm{res}_k\Bigl(\frac{-in\pi}{|qE|}\Bigr)\,.
\end{equation}
We do not include the pole at $T=0$ since it has nothing to do with effects of electromagnetic fields or the axial gauge. To evaluate the residue, notice that the derivative $\partial^k_T$ acting on the hyperbolic cotangent function gives $(k+1)$-th order poles as the highest-order pole of the integrand. Thus, after application of the Leibniz rule to the derivative operators one may find
\begin{align}
    \mathrm{res}_k&\Bigl(\frac{-in\pi}{|qE|}\Bigr)=\frac{d_k}{k!}\sum_{r=0}^k \sum^k_{s=0} {}_kC_r\, {}_k C_s \lim_{T\rightarrow -in\pi /|qE|}\nonumber \\
    &\times \partial^{k-s}_T \Bigl\{ T^{-3/2}e^{-i(m^2-\omega_5^2)T}T^{2k}\partial_T^{k-r}\sqrt{T^{-1}}      \Bigr\}\nonumber \\
    &\times \partial_T^s\Bigl\{  \Bigl(T+\frac{in\pi}{|qE|}  \Bigr)^{k+1}\partial_T^r\coth(|qE|T)    \Bigr\}\,,
\end{align}
where $d_k=(-1)^k(2\omega_5)^{2k} (-i)^k/[(2k)!]$ 
and $_aC_b$ is the binomial coefficient. A key observation here is that only the $\mathcal{O}((T+in\pi/|qE|)^{-1})$ term in a power series expansion of the hyperbolic cotangent term will ultimately remain after taking the limit. Therefore, under the limit and differential operators we may take that $\coth(|qE|T)\sim |qE|^{-1}(T+in\pi/|qE|)^{-1}$. Then one can find that
\begin{align}
    \mathrm{res}_k&\Bigl(\frac{-in\pi}{|qE|}\Bigr)=\frac{1}{|qE|}\sum_{r=0}^k d_k \frac{(-1)^rr!(k-r)!}{k!}\lim_{T\rightarrow -in\pi /|qE|} \nonumber \\
    &\times {}_kC_r\,{}_kC_{k-r} \partial^r_T\Bigl\{  T^{-3/2} e^{-i(m^2-\omega_5^2)T}T^{2k} \partial_T^{k-r}\sqrt{T^{-1}}  \Bigr\}\,.
\end{align}
Then taking the innermost partial derivative, applying the Leibniz rule once more, and taking the limit one can find
\begin{align}
    \mathrm{res}_k&\Bigl(\frac{-in\pi}{|qE|}\Bigr)=\frac{e^{-\frac{(m^2-\omega_5^2)n\pi}{|qE|}}}{|qE|}\sum_{r=0}^k\sum_{l=0}^r {}_kC_r\,{}_kC_{k-r}\,{}_rC_l\nonumber\\
    &\times d_k\frac{(-1)^rr!(k-r)!}{k!}\frac{(-1)^{k-r}(2k-2r-1)!!}{2^{k-r}} \nonumber \\
    &\times [-i(m^2-\omega_5^2)]^{r-l}\beta_{k+r}^l\Bigl( \frac{-in\pi}{|qE|} \Bigr)^{k+r-2-l}\,,
\end{align}
where 
\begin{equation}\label{eq:beta}
    \beta_q^l=
    \begin{cases}
      \frac{(q-2)!}{(q-2-l)!} & \hphantom{\text{for}}\ q-2\geq q-2-l\geq 0 \\
      0 & \text{for}\ q-2 \geq 0\geq q-2-l \\
      (-1)^l \frac{|q-1-l|!}{|q-1|!} & \hphantom{\text{for}}\ 0>q-2 \geq q-2-l
    \end{cases}
    \,.
\end{equation}
Finally after some manipulations we arrive at
\begin{align}\label{eq:res_final}
    \mathrm{res}_k&\Bigl(\frac{-in\pi}{|qE|}\Bigr)=-\frac{1}{|qE|} e^{-\frac{(m^2-\omega_5^2)n\pi}{|qE|}} \sum_{r=0}^k\sum_{l=0}^r \alpha_k^{r,l}\nonumber \\
    &\times \omega_5^{2k} (m^2-\omega_5^2)^{r-l} \Bigl( \frac{n\pi}{|qE|} \Bigr)^{k+r-2-l}\,,
\end{align}
with the coefficient
\begin{equation}\label{eq:alpha}
    \alpha_k^{r,l}=(-1)^{k+r-l} {}_kC_{k-r}\,{}_rC_l \frac{2^{k+r}}{(2k)!}\beta^l_{k+r}(2k-2r-1)!!\,.
\end{equation}
Applying Eq.~\eqref{eq:res_final} with Eqs.~\eqref{eq:beta} and~\eqref{eq:alpha} to Eq.~\eqref{eq:Im_sum_poles} leads to an exact in summation form of the imaginary part of the effective action. 

Let us consider the $\omega_5\rightarrow m$ limit. While such a scenario may be difficult to achieve in most physical systems, including QED, in heavy ion collision experiments recent measurements of global spin polarization of $\Lambda$ hyperons suggest a large angular momentum is produced in the collision event and converted to a quark gluon plasma vortical field whose scale is around $2-20$ MeV~\cite{STAR:2017ckg}; this is comparable to the up and down quark mass of $m\sim 1.8-5.3$ MeV. While we treat QED here, extensions and applicability for quantum chromodynamics hold in Abelian-projected fields. Even so, let us stress that we do not treat the finite temperature and chemical potential that are present in the quark gluon plasma, which may affect our arguments on vorticity and the Schwinger effect. Then for our setup of QED plus a constant axial gauge, we find for $\omega_5\rightarrow m$ that remarkably, the exponential suppression is reduced by a nonzero $\omega_5$ as $\exp[-(m^2-\omega_5^2)n\pi/|qE|]$ in Eq.~(\ref{eq:res_final}). 
When $\omega_5 \to m$, the exponential suppression completely goes away, and the magnitude of the residue is determined by the polynomial at $r=l$ as 
\begin{align}\label{eq:res_final-limit}
    \mathrm{res}_k&\Bigl(\frac{-in\pi}{|qE|}\Bigr)
    = - \sum_{r=0}^k \alpha_k^{r,r}
    \omega_5^{2k} \Bigl( \frac{n\pi}{|qE|} \Bigr)^{k-2}\,.
\end{align}

\begin{figure}
  \includegraphics[width=0.8\columnwidth]{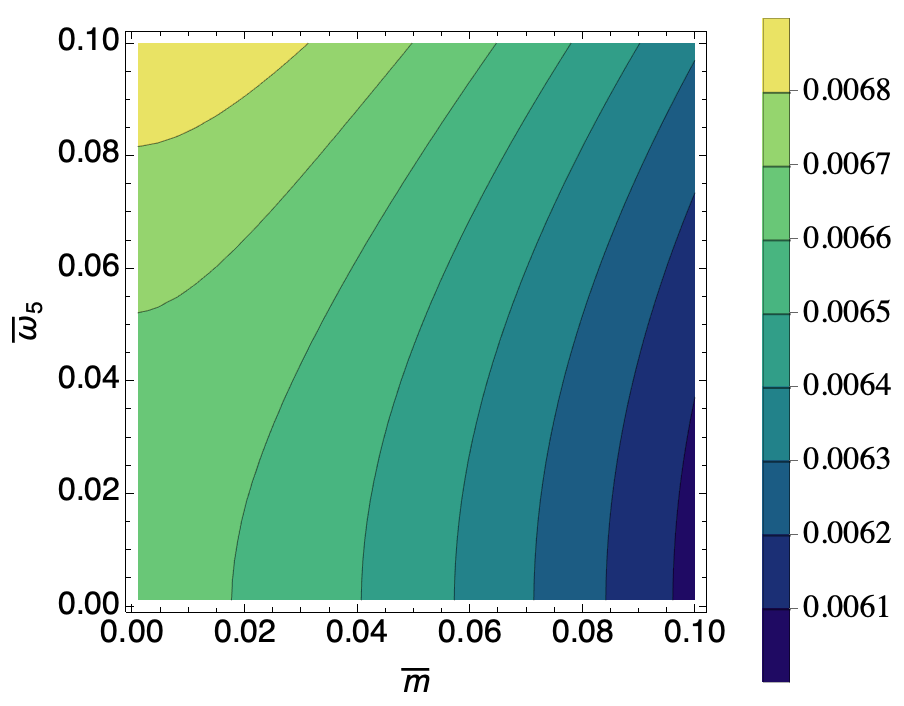}
  \caption{Imaginary part of the effective action,  $\mathrm{Im}\Gamma(E,\omega_{5})$, to $\mathcal{O}(\omega_5^4)$ as a function of normalized mass, $\bar{m}$ and spatial axial gauge, $\bar{\omega}_5$ as given in Eq.~\eqref{eq:imag_full}.  Notice for constant $\bar{m}$, there is an enhancement of pair production for increasing $\bar{\omega}_5$. Also lines of an approximately constant values of the imaginary part are seen for linear relationships between $\bar{m}$ and $\bar{\omega}_5$. }
  \label{fig:axial_pair_production}
\end{figure}

A resummation to a simple form is not achievable with arbitrary parameters; therefore, as before let us examine a perturbative to $\mathcal{O}(\omega_5^4)$ expression. Let us emphasize that this perturbative scheme is about powers of the dimensionless $\omega_5/\sqrt{|qE|}$, and thus here $\omega_5\ll \sqrt{|qE|}$. This in turn implies that one must also have $m \ll \sqrt{|qE|}$ for strong fields or small mass. We find in the perturbative scheme
\begin{align}\label{eq:imag_full}
&\mathrm{Im}\Gamma(E,\omega_{5})=\mathrm{Im}\Gamma(E,0)+\frac{\mathcal{TL}^{3}}{24\pi}\frac{m^{2}\omega_{5}^{2}}{|qE|^{2}}\sinh^{-2}\Bigl(\frac{\pi m^{2}}{2|qE|}\Bigr)\nonumber \\
&\times \Bigl\{3|qE|^{2}e^{-\frac{\pi m^{2}}{2|qE|}}\sinh\Bigl(\frac{\pi m^{2}}{2|qE|}\Bigr)-\pi|qE|\omega_{5}^{2}\nonumber \\
&\quad+\frac{\pi^{2}m^{2}\omega_{5}^{2}}{2}\coth\Bigl(\frac{\pi m^{2}}{2|qE|}\Bigr)\Bigr\}\,.
\end{align}
Here
\begin{equation}
\mathrm{Im}\Gamma(E,0)=\frac{\mathcal{TL}^{3}|qE|^{2}}{8\pi^{3}}\text{Li}_{2}(e^{-\frac{m^{2}\pi}{|qE|}})
\end{equation}
is the usual imaginary part of the effective action for an electric field with no axial gauge coupling. 
Let us also write the expression for the leading order pole contribution truncating the above to an evaluation of the residue for only $T=-i/|qE|$, leading to a correction to the familiar expression of
\begin{align}\label{eq:eff_action_small}
\mathrm{Im}\Gamma(E,\omega_{5})&=\frac{\mathcal{TL}^{3}}{24\pi^{3}|qE|^{2}}e^{-\frac{\pi m^{2}}{|qE|}}\bigl(3|qE|^{4}+6\pi^{2}m^{2}|qE|^{2}\omega_{5}^{2}\nonumber \\
&+[2\pi^{4}m^{4}-4\pi^{3}m^{2}|qE|]\omega_{5}^{4}\bigr)\,.
\end{align}
One can clearly see in the above that contributions to the effective action from the spatial axial gauge are coupled to the mass.  Furthermore, since one has the characteristic exponential suppression, one finds that the enhancement to pair production is pronounced for smaller masses. We keep the exponential expression of quadratic mass suppression, but let us remind the reader that we treat a small mass or large electric field in this perturbative scheme. To show the enhancement see Fig.~\ref{fig:axial_pair_production},  which has been produced for small dimensionless spatial axial gauge and mass
\begin{equation}
    \bar{\omega}_5 =\frac{\omega_5}{\sqrt{|qE|}}\,,\quad \bar{m}=\frac{m}{\sqrt{|qE|}}\,,
\end{equation} 
in accordance with the perturbative scheme at all poles using Eq.~\eqref{eq:imag_full} (let us stress that such small $\bar{m}$ are, however, beyond current capabilities of high-powered lasers to date).  
In Fig.~\ref{fig:axial_pair_production},  one can notice that indeed pair production is enhanced due to $\omega_5$. Moreover, notice as well that lines of constant $\mathrm{Im}\Gamma(E,\omega_{5})$ exist for approximately linear relationship between $\bar{m}$ and $\bar{\omega}_5$ indicating that for a given electric field the same pair production threshold can be met by substituting the mass for the spatial axial gauge.  Or in other words, the spatial axial gauge serves to shift the mass, which is shown here for small $\bar{m}$ and $\bar{\omega}_5$, as visible in Eq.~\eqref{eq:res_final}.  
One may still find a noticeable enhancement of pair production at higher $\bar{\omega}_5$;   
however, the enhancement appears to a polynomial degree, not exponential, in the above perturbative expansion. 
It is therefore important to further discuss the non-perturbative effect on the exponential found in Eq.~(\ref{eq:res_final}). For this purpose we use the semi-classical worldline instanton approach in the next section. 
Doing so will also furthermore provide us with physical insight into the enhancement process.

Before introducing the worldline instanton approach, let us determine the real part of the effective action. Let us proceed as before and write Eq.~\eqref{eq:eff_spatial_diff} to $\mathcal{O}(\omega_5^4)$ for $\textrm{c}_E\equiv\coth(|qE|T)$ as
\begin{align}\label{eq:eff_E_final}
&\Gamma(E,\omega_{5})\approx\frac{\mathcal{TL}^{3}|qE|}{8\pi^{2}}\int_{0}^{\infty}dT \Bigl\{\frac{\textrm{c}_{E}}{T^2}+2i\omega_{5}^{2}|qE|(1-\textrm{c}_{E}^{2}) \nonumber \\
&-\frac{4}{3}\omega_{5}^{4}|qE|T(1-\textrm{c}_{E}^{2})(1-|qE|\textrm{c}_{E}T)\Bigr\}e^{-im^{2}T}\,.
\end{align}
Also like in the previous case with a magnetic field, one can introduce a cutoff in Schwinger propertime at a QED scale of $\Lambda^{-2}$,  which is applied after a deformation of the propertime contour to the imaginary axis, and a change of variables, $T\rightarrow-iT$.  Furthermore, we restrict our attention to the case of weak electric fields to $\mathcal{O}(|qE|^4)$,
and then one can find for the real part of Eq.~\eqref{eq:eff_E_final}
\begin{align}
\textrm{Re}\Gamma(E,\omega_{5})&=\frac{\mathcal{TL}^{3}}{8\pi^{2}}\int_{\Lambda^{-2}}^{\infty}\frac{dT}{e^{m^{2}T}}\Bigl\{-\frac{1}{T^{3}}+\frac{|qE|^{2}}{3T}+\frac{|qE|^{4}T}{45}\notag\\
&+2\omega_{5}^{2}\Bigl(\frac{1}{T^{2}}+\frac{1}{3}|qE|^{2}+\frac{1}{15}|qE|^{4}T^{2}\Bigr)\notag\\
&+\frac{4}{9}\omega_{5}^{4}|qE|^{2}T\Bigl(1+\frac{2}{5}|qE|^{2}T^{2}\Bigr)\Bigr\}\,.
\end{align}
Then using the integrals in Eqs.~\eqref{eq:int_1}-\eqref{eq:int_4} one can find for $\Gamma_{\text{reg.}}[A,A_5]=\Gamma[A,A_5]-\Gamma[0,0]$ that
\begin{align}
\mathrm{Re}\Gamma&(E,\omega_{5})_\text{reg.}=\frac{\mathcal{TL}^{3}}{8\pi^{2}}\Bigl\{-\frac{|qE|^{2}}{3}\ln\bigl(\frac{m^{2}}{\Lambda^{2}}\bigr)+\frac{1}{45}\frac{|qE|^{4}}{m^{4}}\notag\\
&+2\omega_{5}^{2}\Bigl(\Lambda^{2}+m^{2}\ln\bigl(\frac{m^{2}}{\Lambda^{2}}\bigr)+\frac{1}{3}\frac{|qE|^{2}}{m^{2}}\Bigr)\Bigr\}\,,
\end{align}
where we have as before kept only terms quartic in either $|qE|$ or $\omega_5$, omitting for example terms that go like $\mathcal{O}(\omega_5^4|qE|^2)$. The key difference we see in this expression compared to Eq.~\eqref{eq:eff_real_magnetic} is that whereas the chiral chemical potential  decreased the system energy, we see here that the spatial axial gauge increases the energy for the dominant $\Lambda^2$ factor indicating an instability.  This is to be expected as we have observed above that indeed an instability is present in the form of an imaginary part of the effective action, whereby the spatial axial gauge further increases the imaginary part. $\Gamma[0,0]$ here is the same as it was for Eq.~\eqref{eq:eff_free}. It is interesting to investigate non-Abelian extensions (see, e.g., Refs.~\cite{Batalin:1976uv, *Matinyan:1976mp, *Savvidy:1977as, *Yildiz:1979vv, *Dittrich:1983ej, *Casher:1978wy, *Ozaki:2015yja, *Bali:2013esa}) that may capture some aspects of intertwined dynamics with nonzero topological configurations of the gauge field.  

As before in the case with the magnetic field and chiral chemical potential, one may define a renormalization scheme using $\overline{{\rm MS}}$. We have the similar divergences as well, i.e., a divergence at $\mathcal{O}(T^{-1})$ associated with the electromagnetic coupling constant and the electric field, and a divergence at $\mathcal{O}(T^{-2})$ associated with the spatial axial gauge field. Let us denote for our bare quantities $q_0$ and $E_0$, and our renormalized physical quantities as $q=q_0\sqrt{Z_3}$ and $E=E_0/\sqrt{Z_3}$ such that $qE=q_0E_0$. Then, the Euler-Heisenberg Lagrangian under the $\overline{{\rm MS}}$ scheme becomes
\begin{align}
    \mathbb{L}_\text{EH}&=\frac{E^2}{2}+ \frac{1}{8\pi^{2}}\int_{0}^{\infty}\frac{dT}{e^{im^{2}T}} \Bigl\{\frac{|qE|c_E}{T^2}-\frac{|qE|^2}{3T}-\frac{1}{T^3}\notag\\
    &+2i\omega_{5}^{2}\Bigl[|qE|^2(1-c_{E}^2)+\frac{1}{T^2}\Bigr]\Bigr\}\,,
\end{align}
with renormalization factor
\begin{equation}
Z_3=1+\frac{q_0^2}{12\pi^2}\int^{\infty}_{\Lambda^{-2}}\frac{dT}{T}e^{-im^2T}\Big(1-\frac{6i\omega_5^2}{(q_0B_0)^2T}\Big)\,.
\end{equation}
We emphasize as before, that a similar Euler-Heisenberg augmented Lagrangian could be had with a subtraction of the vacuum energy with $\bar \Gamma_{\text{reg.}}[A,A_5]=\Gamma[A,A_5]-\Gamma[0,A_5]$.

\section{Worldline Instanton Method}
\label{sec:worldline}

Above it was demonstrated that with a spatial axial gauge background the threshold for Schwinger pair production is reduced. 
Here we analyze its non-perturbative structure using the semi-classical worldline instanton method~\cite{AFFLECK1982509,*PhysRevD.72.105004}.  This has the added benefit of providing physical insight into the enhancement process.  In contrast to conventional approaches used in applying the worldline instanton method, here with the addition of the axial gauge, the quadratic form of the Dirac operator, Eq.~\eqref{eq:quad_form}, is unnecessarily complicated for our purposes hindering usage of worldine formalism in configuration space. Therefore, we elect to use a linear in Dirac operator phase space worldline representation~\cite{Migdal:1986pz,*doi:10.1142/7305,*Fosco:2003rr}.

Beginning with the effective action in logarithmic form,  Eq.~\eqref{eq:eff_action_def},  one can construct a phase space path integral representation with spinor degrees of freedom captured in coherent state form through the use of BZ spinor formalism~\cite{PhysRevLett.52.2009,Barut:1988ud,*Barut:1984vnk}.  Since the worldline instanton method relies on the determination of classical solutions, usage of a coherent state is valuable in that it provides one with a well-defined scalar action as opposed to a path-ordered and matrix weighted action without.  For application of BZ spinors, we use the approach followed in Ref.~\cite{PhysRevD.105.116014}.  First, let us write down the effective action in its path integral form with path-ordering and without coherent states as
\begin{align} \label{eq:path_int_before}
   \Gamma[A,A_5]&=i\mathrm{tr}\int_{0}^{\infty}\frac{dT}{T}\oint\mathcal{D}x
   \int\frac{\mathcal{D}p}{2\pi}\,\mathcal{P} e^{i S_{\textrm{W}}}\,,\\
S_{\textrm{W}}&\coloneqq\int_{0}^{T}d\tau[-m-p_{\mu}\dot{x}^{\mu}
   -q A_{\mu}\dot{x}^{\mu}+\slashed{\Pi}_5]\,,\\
\slashed{\Pi}_5&=\slashed{p}+\gamma_5\slashed{A}_5\,.
\end{align}
Here $\oint\mathcal{D}x=\int dx'\int \mathcal{D}x$ denotes a periodic path integral with boundary conditions $x(0)=x(T)=x'$. Also, note that we have left implicit the small imaginary component resolving the causal nature of the propagator and also providing convergence in the IR into the mass such that $m\rightarrow m-i\epsilon$. Then, one need only split the path-ordered expression into a product of infinitesimal propertime elements, inserting into each the resolution of identity of the BZ spinors given as
\begin{equation}
   \int d\Omega_{z}\,z_{a}\bar{z}_{b}\coloneqq\mathcal{Z}_{0}^{-1}\partial_{\bar{\eta}_{a}} 
   \partial_{\eta_{b}}\mathcal{Z}_{\eta}\Big|_{\eta=0}=\delta_{ab}\,,
   \label{eq:identity}
\end{equation}
where $\mathcal{Z}_{\eta}=\int dzd\bar{z}\exp(-\bar{z}z+\bar{z}\eta+\bar{\eta}z)
   =\mathcal{Z}_{0}\exp(\bar{\eta}\eta)$ with $\mathcal{Z}_{0}=\int d\bar{z}dz\,\exp(-\bar{z}z)=\pi^{4}$~\cite{PhysRevD.105.116014}.  After summing the infinitesimal elements,  and also taking a change of variables such that $\tau\rightarrow T\tau$, one can finally find for the effective action the following form:
\begin{align}
&\Gamma[A,A_{5}]=i\int_{0}^{\infty}\frac{dT}{T}\oint\mathcal{D}x\mathcal{D}\Omega_z \int\frac{\mathcal{D}p}{(2\pi)^{4}}\,e^{iS_{\textrm{W}}'}\label{eq:path_integral}\\
&S_{\textrm{W}}'\coloneqq \int_{0}^{1}d\tau[-mT-p_{\mu}\dot{x}^{\mu}-qA_{\mu}\dot{x}^{\mu}+i\bar{z}\dot{z}+T\bar{z}\slashed{\Pi}_5z]\,;\label{eq:S_w}
\end{align}
the ``dots'' now represent total derivatives with respective to the integration variable. 

In configuration space and in Euclidean spacetime, worldline instantons represent periodic classical solutions found at stationary points in $x^\mu$ and propertime $T$,  which for simple field configurations are expressible for real coordinates thanks to an inverted time after Wick rotation.  Here in Minkowski spacetime we have complex worldline instantons~\cite{PhysRevD.84.125023}; this can be seen in the fact that the stationary points (poles in the Euler-Heisenberg Lagrangian) lie on the imaginary propertime axis.  An additional unique feature is that we are in phase-space and worldline instantons can be found at stationary points in $p^\mu$, $x^\mu$, and $T$, whose classical equations of motion of Eq.~\eqref{eq:S_w} are
\begin{align}
\dot{x}_{\mu} & =T\bar{z}\gamma_{\mu}z\,,\label{eq:eom_x}\\
\dot{p}_{\mu} & =qF_{\mu\nu}\dot{x}^{\nu}\,,\label{eq:eom_p}\\
m & =\bar{z}\slashed{\Pi}_5z\,.\label{eq:eom_z}
\end{align}
Here, we assume that the axial gauge $A_5^\mu$ is independent of $x^\mu$, 
but can otherwise take a general configuration. Let us also emphasize that we do not directly treat the configurations presented in Eqs.~\eqref{eq:mag_field}-\eqref{eq:elec_field}.
Propertime in Eq.~\eqref{eq:S_w} acts as a Lagrange multiplier, sending the particle on-shell as can be seen in Eq.~\eqref{eq:eom_z} above.  Let us remark that for non-Abelian systems coherent state variables extend the phase-space leading to Wong's equations~\cite{Wong:1970fu} instead of just the Lorentz force.  And it is instructive that here, too, stationary points should lie in the entire phase space, including $\bar{z}$ and $z$.  However we already have the mass-shell constraint in Eq.~\eqref{eq:eom_z} that will provide solutions in $\bar{z}$ and $z$.  We must look for eigenvectors of a Dirac equation in the presence of an axial gauge background. There are four independent solutions, 
\begin{equation}
z\in u_{i}\,,v_{i}\quad \text{for}\quad i=1,2
\end{equation}
that are closely analogous to ordinary Dirac spinors that satisfy
\begin{equation}\label{eq:P_5}
\slashed{\Pi}_5z=\lambda z\,,
\end{equation}
for eigenvalue, $\lambda$ to be determined.  In this way we find worldline instantons can be found for solutions in Eqs.~\eqref{eq:eom_x}-\eqref{eq:eom_z} that lead to an imaginary part of the effective action; and according to Eq.~\eqref{eq:path_integral},  we have the periodic boundary conditions: $x^\mu(0)=x^\mu(1)$, $\bar{z}(0)=\bar{z}(1)$, and $z(0)=z(1)$. Let us remark that there is an ambiguity in the ordering of the evaluation of the propertime integral, $T$, and those in $z$ and $\bar{z}$; we further remark on their differences and choice of selection leading to Eqs.~\eqref{eq:eom_x}-\eqref{eq:eom_z} in  Appendix A\ref{sec:appendix}. A feature of the worldline instanton method is a weak electric field or large mass approximation. We also further show in the Appendix A that our usage of Eqs.~\eqref{eq:eom_x}-\eqref{eq:eom_z} amounts to the adiabatic theorem used in finding the Berry's phase~\cite{doi:10.1098/rspa.1984.0023}, and furthermore that the theorem in fact implies the weak electric field or large mass approximation is applicable to our calculation as well.

Since we wish to extract the leading order exponential suppression, additionally, we neglect prefactor fluctuations about the worldline instantons~\cite{PhysRevD.73.065028}.  One can then find for worldline instantons with winding number $n$ that
\begin{equation}
\textrm{Im}\Gamma[A,A_{5}] \approx\sum_{z=u_{\pm},v_{\pm}}\sum_{n=1}^\infty e^{i S_{z,n}}\,,
\end{equation}
where $S_{z,n}$ is the worldline action, Eq.~\eqref{eq:S_w}, evaluated for the classical worldline instanton solution. Since $T$ is linear in Eq.~\eqref{eq:S_w} and moreover since Eq.~\eqref{eq:eom_z} holds,  there is no fluctuation term and its evaluation about a given stationary point only introduces a numerical prefactor. 

Let us proceed with the evaluation of the equations of motion.  We begin by determining the eigensystem of $\slashed{\Pi}_5$ shown in Eq.~\eqref{eq:P_5}.  Its quadratic and quartic forms read
\begin{align}
\slashed{\Pi}_5^{2}&=p^{2}-A_{5}^{2}+2i\gamma_{5}\sigma_{\mu\nu}p^{\mu}A_{5}^{\nu}\,,\\
\slashed{\Pi}_5^{4}&=2(p^{2}-A_{5}^{2})\slashed{\Pi}_5^{2}-4[p^{2}A_{5}^{2}-(p_{\mu}A_{5}^{\mu})^{2}]-(p^{2}-A_{5}^{2})^{2}\,,\label{eq:P_5^4}
\end{align}
where we have made use of the fact that $(\sigma_{\mu\nu}p^{\mu}A_{5}^{\nu})^{2}=p^{2}A_{5}^{2}-(p_{\mu}A_{5}^{\mu})^{2}$.  We have also assumed implicit identity elements where appropriate for the (3+1)-dimensional Clifford group. We can see that this is nothing more than the characteristic equation leading to the eigenvalues of $\slashed{\Pi}_5$ according to Cayley-Hamilton's theorem with $\slashed{\Pi}_5\rightarrow\lambda$. Therefore we find the eigenvalues of $\slashed{\Pi}_5$ are
\begin{equation}
\lambda_{\pm'\pm''} =\pm'\sqrt{p^{2}-A_{5}^{2}\pm''2\sqrt{-p^{2}A_{5}^{2}+(p_{\mu}A_{5}^{\mu})^{2}}}\,.
\end{equation}
One can compare the above to operator eigenvalues that were determined for the magnetic field case in Eq.~\eqref{eq:magnetic_eigenvalues}, and the electric field case in Eq.~\eqref{eq:electric_eigenvalues}.  However, we caution that the c-number momentum here represents the kinetic momentum, whereas above it represents a canonical momentum.  Nevertheless, one can see the similar structure in $A_5^\mu$ present for all.  In the absence of $A_5^\mu$ it can be seen that the eigenvalues would describe degenerate $\pm \sqrt{p^2}$.  A novelty here is that with the presence of $A_5^\mu$ the degeneracy is lifted, which we will later show impacts pair production.

Then to determine the eigenvectors of $\slashed{\Pi}_5$,  let us next construct projection operators from the above quartic equation as
\begin{align}
P_{+} & =\frac{-i\gamma_{5}\sigma_{\mu\nu}p^{\mu}A_{5}^{\nu}}{2\sqrt{-p^{2}A_{5}^{2}+(p_{\mu}A_{5}^{\mu})^{2}}}+\frac{1}{2}\,,\\
P_{-} & =\frac{i\gamma_{5}\sigma_{\mu\nu}p^{\mu}A_{5}^{\nu}}{2\sqrt{-p^{2}A_{5}^{2}+(p_{\mu}A_{5}^{\mu})^{2}}}+\frac{1}{2}\,.
\end{align}
One can confirm that $P_{+}P_{-}=0$, that $P_{\pm}^{2}=P_{\pm}$, and that $P_{+}+P_{-}=1$. The projection operators have the effect of taking $\slashed{\Pi}_5^{2}P_{\pm''}=\lambda_{\mp' \mp''}^{2}P_{\pm''}$. 
Finally let us construct orthonormalized eigenvectors using the projection operators.  For the following eigenvectors corresponding to eigenvalues:
\begin{align}
\slashed{\Pi}_5u_{\pm}&=\lambda_{+\pm}u_{\pm}\,,\quad\bar{u}_{\pm}\slashed{\Pi}_5=\bar{u}_{\pm}\lambda_{+\pm}\,,\label{eq:P_5andu}\\
\slashed{\Pi}_5v_{\pm}&=\lambda_{-\pm}v_{\pm}\,,\quad\bar{v}_{\pm}\slashed{\Pi}_5=\bar{v}_{\pm}\lambda_{-\pm}\,,\label{eq:P_5andv}
\end{align}
one can determine that
\begin{align}
u_{\pm}&=N_{\pm}^{u}(\slashed{\Pi}_5-\lambda_{-\pm})P_{\mp}\xi_{\pm}\,,\\
\bar{u}_{\pm}&\equiv N_{\pm}^{u}\xi_{\pm}^{T}\gamma^{0}P_{\mp}(\slashed{\Pi}_5-\lambda_{-\pm})\,,\\
v_{\pm}&=N_{\pm}^{v}(\slashed{\Pi}_5-\lambda_{+\pm})P_{\mp}\eta_{\pm}\,,\\
\bar{v}_{\pm}&\equiv N_{\pm}^{v}\eta_{\pm}^{T}\gamma^{0}P_{\mp}(\slashed{\Pi}_5-\lambda_{+\pm})\,.
\end{align}
where $\gamma^{0}\xi_{i}=\xi_{i}$, $\gamma^{0}\eta_{i}=-\eta_{i}$, and $\xi_{i}^{T}\xi_{j}=\eta_{i}^{T}\eta_{j}=\delta_{ij}$ with other contractions vanishing.  Note that in general $\bar{u}_\pm= u_{\pm}^{\dagger}\gamma^{0}$ need not be true and likewise for $\bar{v}_\pm$. This is because we will find worldline instantons in Minkowski space are in general complex.  We caution that for strictly real solutions outside of our treatment of instantons here, for constant but arbitrary $A_5^\mu$,  one could find spacelike $p^\mu$ pointing to a potential stability issue; see related arguments in, e.g., ~\cite{PhysRevD.63.065008}.  Let us also remark that the above reduce to ordinary Dirac spinors in the Weyl representation upon carefully taking the $A_{5}^{\mu}\rightarrow\pm0$ limit in a symmetric way. We also take that $\xi_{1}^{T}=[1,0,1,0]$ , $\xi_{2}^{T}=[0,1,0,1]$, $\eta_{1}^{T}=[1,0,-1,0]$, and $\eta_{2}^{T}=[0,1,0,-1]$. The eigenvectors are normalized such that $\bar{u}_{\pm}u_{\pm}=-\bar{v}_{\pm}v_{\pm}=1$; this entails that
\begin{align}\label{eq:normalization}
(N_{\pm}^{u})^2&=[-2\lambda_{-\pm}\xi_{\pm}^{T}(\slashed{\Pi}_5-\lambda_{-\pm})P_{\mp}\xi_{\pm}]^{-1}\,,\\
(N_{\pm}^{v})^2&=[-2\lambda_{+\pm}\eta_{\pm}^{T}(\slashed{\Pi}_5-\lambda_{+\pm})P_{\mp}\eta_{\pm}]^{-1}\,.
\end{align}

Next, taking a partial derivative with respect to $p^\mu$--we denote as $\partial_\mu^p$--in Eqs.~\eqref{eq:P_5andu}-\eqref{eq:P_5andv} one can determine that the velocities, according to Eq.~\eqref{eq:eom_x}, are
\begin{equation}\label{eq:velocities}
\bar{u}_{\pm}\gamma_{\mu}u_{\pm}=\partial_{\mu}^p\lambda_{+\pm}\,,\quad
\bar{v}_{\pm}\gamma_{\mu}v_{\pm}=-\partial_{\mu}^p\lambda_{-\pm}\,.
\end{equation}
Note that we acquire a minus sign in $v_\pm$ due to their normalization. 
\begin{equation}
\partial_{\mu}^{p}\lambda_{\pm'\pm''}=\frac{1}{\lambda_{\pm'\pm''}}\Bigl\{p_{\mu}\pm''\frac{A_{5\mu}(p_{\nu}A_{5}^{\nu})-p_{\mu}A_{5}^{2}}{\sqrt{-p^{2}A_{5}^{2}+(p_{\mu}A_{5}^{\mu})^{2}}}\Bigr\}\,.
\end{equation}
At this point we must emphasize that there are four separate solutions, and with each one in general different values of $p_\mu$ and $x_\mu$ such that the three equations of motion, Eqs.~\eqref{eq:eom_x}-\eqref{eq:eom_z}, can be satisfied.  To reduce cumbersome notation, we leave this distinction implicit in $p_\mu$ and $x_\mu$. This fact is readily apparent for Eq.~\eqref{eq:eom_z} in which the on-shell constraint is taken; the four distinct solutions follow from
\begin{equation}\label{eq:m_eigenvalue}
\bar{u}_{\pm}\slashed{\Pi}_5u_{\pm}=\lambda_{+\pm}=m\,,\quad\bar{v}_{\pm}\slashed{\Pi}_5v_{\pm}=-\lambda_{-\pm}=m\,.
\end{equation}
The above then entails that the following conditions be met for each set of solutions:
\begin{align}
u_{+}, v_+:\;&\sqrt{-p^{2}A_{5}^{2}+(p_{\mu}A_{5}^{\mu})^{2}}=\frac{1}{2}[m^{2}-p^{2}+A_{5}^{2}] \label{eq:+_constraint}\\
u_{-},v_-:\;&\sqrt{-p^{2}A_{5}^{2}+(p_{\mu}A_{5}^{\mu})^{2}}=\frac{1}{2}[p^{2}-m^{2}-A_{5}^{2}]\,.
\label{eq:-_constraint}
\end{align}

With application of the above constraint, we can modify Eq.~\eqref{eq:velocities}; for example here we turn our attention to just the $u_+$ solution, whose velocity may now be written as
\begin{equation}\label{eq:u_+_velocity}
\bar{u}_{+}\gamma_{\mu}u_{+} \eqqcolon\frac{1}{m^{*}}(p_{\mu}+\mathcal{C}_{\mu})=\frac{1}{T}\dot{x}_{\mu}\,,
\end{equation}
where we see the momentum has been modified with an augmented mass term as well as a shift:
\begin{align}
m^{*}&=m\frac{m^{2}-p^{2}+A_{5}^{2}}{m^{2}-p^{2}-A_{5}^{2}}\,,\label{eq:eff_mass}\\
\mathcal{C}_\mu&=\frac{2p_{\nu}A_{5}^{\nu}A_{5\mu}}{m^{2}-p^{2}-A_{5}^{2}}\,.
\end{align}
One can find similar expressions for $u_-$ and $v_\pm$.  With the addition of an axial gauge field, we will find the usual constraint $p^2=m^2$ no longer applies.  The Lorentz force equation, Eq.~\eqref{eq:eom_p}, now becomes for $u_+$
\begin{equation}\label{eq:dotpmu}
\dot{p}_{\mu}=\frac{qT}{m^{*}}F_{\mu\nu}(p^{\nu}+\mathcal{C}^{\nu})\,.
\end{equation}
According to the constraint given in Eq.~\eqref{eq:+_constraint},  we can determine that the invariants $p^2$ and $A_5^\mu p_\mu$ must be constants of motion and independent of propertime. 
Then we can evaluate the Lorentz force equation as $p(\tau)=\exp[(qT/m^*)F\tau]k-\mathcal{C}$ for $k=p(0)+ \mathcal{C}$, 
or rather
\begin{align}
p(\tau)&=f_{E}(\tau)P_{E}k+f_{B}(\tau)P_{B}k-\mathcal{C}\,,\label{eq:p_soln}\\
f_E(\tau)&=\cosh\bigl(\frac{q\lambda_{E}T\tau}{m^{*}}\bigr)+\frac{F}{\lambda_{E}}\sinh\bigl(\frac{q\lambda_{E}T\tau}{m^{*}}\bigr)\,,\label{eq:f_E}\\
f_B(\tau)&=\cos\bigl(\frac{q\lambda_{B}T\tau}{m^{*}}\bigr)+\frac{F}{\lambda_{B}}\sin\bigl(\frac{q\lambda_{B}T\tau}{m^{*}}\bigr)\,.\label{eq:f_B}
\end{align}
Note that when $ p^2$ and $ A_5^\mu p_\mu$ are constants, so are $C^\mu $ and $m^*$. 
Thus, $k^\mu$ is independent of propertime. Here and where appropriate we use a matrix form for tensors and vectors with Lorentz indices, e.g., $F^\mu_{\; \nu}\equiv F$ and $p^\mu\equiv p$, where contractions are assumed.  
An exact solution to the Lorentz force is possible with the use of the projection operators~\cite{Fradkin_1978}
\begin{equation}
P_{E}=\frac{\lambda_{B}^{2}+F^{2}}{\lambda_{B}^{2}+\lambda_{E}^{2}}\,,\quad P_{B}=\frac{\lambda_{E}^{2}-F^{2}}{\lambda_{B}^{2}+\lambda_{E}^{2}}\,,
\end{equation}
which satisfy $P_E^2=P_E$, $P_B^2=P_B$,  $P_EP_B=0$,  and $P_E+P_B=1$,  and act to decouple the Lorentz force equation.  Note that we use similar conventions as employed in~\cite{Copinger:2022gfz}.  The electric and magnetic field eigenvalue strengths are, respectively,
\begin{align}
\lambda_{E}&=\frac{1}{\sqrt{2}}\sqrt{\sqrt{I_{FF}^{2}+4I_{\tilde{F}F}^{2}}-I_{FF}}\,,\\
\lambda_{B}&=\frac{1}{\sqrt{2}}\sqrt{\sqrt{I_{FF}^{2}+4I_{\tilde{F}F}^{2}}+I_{FF}}\,,
\end{align}
where $I_{\tilde{F}F}=-(1/8)\epsilon_{\mu\nu\alpha\beta}F^{\mu\nu}F^{\alpha\beta}=-(1/4)\widetilde{F}_{\mu\nu}F^{\mu\nu}$ and $I_{FF}=(1/2)F_{\mu\nu}F^{\mu\nu}$ are the Lorentz invariants. Then the coordinate solution follows from Eq.~\eqref{eq:eom_x} as
\begin{equation}\label{eq:x_soln}
x(\tau)=\frac{1}{q}F^{-1}[e^{\frac{qT}{m^{*}}F\tau}-1]k+x(0)\,;
\end{equation}
one can understand the exponential from Eq.~\eqref{eq:p_soln}. Then we find in order to satisfy the periodicity requirement, $x(0)=x(1)$,  using the above we must have that
\begin{equation}
 (f_{E}(1)-1)P_{E}k=0\,,\quad(f_{B}(1)-1)P_{B}k=0\,.
\end{equation}
One may determine a set of stationary points associated with the electric degrees of freedom as we wish to evaluate the effects of pair production; these are located for $f_E(1)=1$ at
\begin{equation}\label{eq:T_u_+}
T=-\frac{2n\pi im^{*}}{q\lambda_{E}}\quad\forall n\in\mathbb{Z}^+\,.
\end{equation}
One may then select trivial magnetic like solutions through the initial condition $P_Bk=0$.

To further search for viable worldline instanton solutions let us examine the constant Lorentz invariant, $A_5^\mu p_\mu$. By taking the propertime total derivative one can find that $p_\mu F^{\mu\nu} A_{5\nu}=0$ must hold, and hence that $A_5^\mu [Ff_E(\tau)P_E]_{\mu\nu}k^\nu=0$ must hold as well. One cannot find a solution in the propertime independent $k^\mu$ for all propertimes. Therefore, one is left with an orthogonality constraint in the instanton momentum and axial gauge field, or rather $A_5^\mu p_\mu=0$; this in turn requires that $\mathcal{C}^\mu=0$.
Since the magnetic degrees of freedom are trivial the constraint implies that
\begin{equation}\label{eq:PE_constraint}
A_{5\mu} P_{E}^{\mu\nu} p_\nu=0\,.
\end{equation}
To further emphasize this point let us take a Lorentz transformation that will effectively diagonalize our field strength tensor as (where we have made use of the matrix form)
\begin{equation}
F_{\;\,\nu}^{\mu}\rightarrow\Lambda^{-1}F\Lambda=\begin{pmatrix} &  &  & \lambda_{E}\\
 &  & \lambda_{B}\\
 & -\lambda_{B}\\
\lambda_{E}
\end{pmatrix}\,,
\end{equation}
and then the projection operators have the simple form
\begin{equation}
P_{E\nu}^\mu = P_{E}=\begin{pmatrix}1\\
 & 0\\
 &  & 0\\
 &  &  & 1
\end{pmatrix}\,,\quad P_{B}=\begin{pmatrix}0\\
 & 1\\
 &  & 1\\
 &  &  & 0
\end{pmatrix}\,.
\end{equation}
Solutions to the momenta follow as
\begin{align}
p^0&=\cosh\Bigl(\frac{q\lambda_{E}T\tau}{m^{*}}\Bigr)k^{0}+\sinh\Bigl(\frac{q\lambda_{E}T\tau}{m^{*}}\Bigr)k^{3}\,,\label{eq:p^0}\\
p^3&=\cosh\Bigl(\frac{q\lambda_{E}T\tau}{m^{*}}\Bigr)k^{3}+\sinh\Bigl(\frac{q\lambda_{E}T\tau}{m^{*}}\Bigr)k^{0}\,.\label{eq:p^3}
\end{align}
Since $P_Bk=0$,  we have that $p^1=p^2=0$. Last, let us assume the axial gauge we are studying is
the one after the unique Lorentz transformation leading to the above, i.e.,  $A^{\mu}\rightarrow(A^{\mu})'=\Lambda_{\;\nu}^{\mu}A^{\nu}$. Then according to the axial gauge as described above, we find that the constraint in Eq.~\eqref{eq:PE_constraint} implies that in order to find worldline instanton solutions we must have $A^0_5 k^0 -A^3_5 k^3=0$ and $A^0_5 k^3 - A^3_5 k^0 =0$. However, for $A_5^{0,3}\neq0$ this constraint would ultimately imply that $k^2=p^2(0)=0$, which we will shortly find would give a trivial worldline instanton solution.
Therefore, we take
$A^0=A^3=0$ with only $A^1$ and $A^2$ contributing.  
This leads us to the following definition for the axial gauge field invariant:
\begin{equation}\label{eq:wi_omega_5}
\omega_5=\sqrt{-A_5^2}\,;
\end{equation}
while such a distinction is not in one-to-one correspondence to the definition used for the spectral decomposition in Eq.~\eqref{eq:elec_field}  (in particular since here we have projected the electric eigenvalue strength, $\lambda_E$, rather than the background electric field),  the above is useful for comparison purposes.  Let us, however, caution that differences do exist for the setups and assumptions here in contrast to those used in Sec.~\ref{sec:eigenvalues}. We will find a similar treatment holds for the $u_-$ and $v_\pm$ solutions as well as leading to Eq.~\eqref{eq:wi_omega_5} for all cases.

Before further pressing on with the $u_+$ solution, let us determine the various Berry phase factors for all $u_\pm$ and $v_\pm$ under the condition $p_\mu A^\mu_5=0$.  Let us begin by expressing the Berry phases as
\begin{equation}
\bar{u}_{\pm}\dot{u}_{\pm}=\frac{N_{\pm}^{u2}}{2}\xi_{\pm}^{\dagger} \mathcal{B}^u_\pm \xi_{\pm}\,,\quad \bar{v}_{\pm}\dot{v}_{\pm}=-\frac{N_{\pm}^{v 2}}{2}\eta_{\pm}^{\dagger}\mathcal{B}^v_\pm \eta_{\pm}\,,
\end{equation}
with arguments entirely expressible in terms of commutation relations in 
\begin{align}
\mathcal{B}^u_\pm &=\Bigl[P_{\mp}(\slashed{\Pi}_5-\lambda_{-\pm}),\frac{d}{d\tau}(\slashed{\Pi}_5-\lambda_{-\pm})P_{\mp}\Bigr]\,,\\
\mathcal{B}^v_\pm &=\Bigl[P_{\mp}(\slashed{\Pi}_5-\lambda_{+\pm}),\frac{d}{d\tau}(\slashed{\Pi}_5-\lambda_{+\pm})P_{\mp}\Bigr]\,.
\end{align}
After some lengthy but straightforward manipulations one can find that the above reduce to
\begin{align}
\mathcal{B}^u_\pm&=\frac{\lambda_{-\pm}}{p^{2}\omega_{5}^{2}}\Bigl\{\omega_{5}^{2}\lambda_{-\pm}-\bigl[\omega_{5}^{2}+\omega_5\sqrt{p^{2}}\bigr]\gamma_{5}\slashed{A}_5 \Bigr\}\slashed{p}\slashed{\dot{p}}\,,\\
\mathcal{B}^v_\pm&=\frac{\lambda_{+\pm}}{p^{2}\omega_{5}^{2}}\Bigl\{\omega_{5}^{2}\lambda_{+\pm}-\bigl[\omega_{5}^{2}+\omega_5\sqrt{p^{2}}\bigr]\gamma_{5}\slashed{A}_5 \Bigr\}\slashed{p}\slashed{\dot{p}}\,.
\end{align}
However, with the application of $\xi_\pm$ and $\eta_\pm$ as given above Eq.~\eqref{eq:normalization}, and the application of worldline instanton solutions in momentum space, Eqs.~\eqref{eq:p^0}-\eqref{eq:p^3} as well as the fact that $A_5^0=A_5^3=0$, one can confirm for the $u_+$ solution,  the Berry phase disappears, $\bar{u}_+\dot{u}_+=0$. In a similar way, once worldline instanton solutions are provided one can also confirm the Berry phases for the remaining spinor solutions also vanish; all together
\begin{equation}
\bar{u}_\pm \dot{u}_\pm = \bar{v}_\pm \dot{v}_\pm=0\,.
\end{equation}
The disappearance of the Berry phase factors for the case of $A^1=A^2\neq0$ occurs from the decoupling nature of the axial gauge degrees of freedom with those of the instantons. Indeed, from a physical standpoint it was determined that a vanishing Berry phase was present in systems with inversion and time-reversal symmetries present~\cite{Nonreciprocal}, here the spatial axial gauge does not break parity, and hence the time-reversal symmetry should also be present. Let us furthermore note that even for the case of no axial gauge the Berry phase term is trivial; one can understand this from the familiar quadratic and in coordinate space expression with spin factor, there the spin factor, and hence spin degrees of freedom, does not affect instanton trajectories.

Returning to the evaluation of the $u_+$ solution,  by invoking Cayley-Hamilton's theorem, Eq.~\eqref{eq:P_5^4},  once more (now under the constraint $A_5^\mu p_\mu=0$),  and evaluating for the momentum in terms of the eigenvalue, we can determine that two solutions for the momentum invariant are possible: $p^2=(m\pm \omega_5)^2$.  We need to find which solution is valid for $u_+$, and therefore we insert both solutions for $p^2$ into the constraint given in Eq.~\eqref{eq:+_constraint}.
Then using the fact that $A_5^2<0$, we can determine that for the $u_+$ solution
\begin{equation}\label{eq:mass_shell}
p^2 = (m-\omega_5)^2\,.
\end{equation}
Furthermore in order to satisfy Eq.~\eqref{eq:+_constraint}, we find that $m\geq \omega_5$ must also hold.
Hence, with the addition of the axial gauge the mass-shell condition has been augmented, in this way using the above, one can find that the effective mass, Eq.~\eqref{eq:eff_mass},  has also been augmented,
\begin{equation}\label{eq:augmented_mass}
m^*=m-\omega_5\,,
\end{equation}
leading to augmented stationary point on the imaginary propertime axis according to Eq.~\eqref{eq:T_u_+}. It is in this way that pair production is enhanced.  Let us now determine the worldline action for the instanton solutions, and to do so let us select a gauge; we use the Fock-Schwinger gauge, i.e., $A_\mu=-(1/2)F_{\mu\nu}x^\nu$. 
Using Eqs.~\eqref{eq:u_+_velocity} and \eqref{eq:dotpmu}, one could obtain $q\int^1_0d\tau A_{\mu}\dot{x}^{\mu}=\int^1_0d\tau \dot{p}_{\mu}x^{\mu}/2=-\int^1_0d\tau p_{\mu}\dot{x}^{\mu}/2$ from the integration by parts and cancellation of surface terms. With Eq.~\eqref{eq:eom_z} and vanishing Berry phases, the remaining term in Eq.~\eqref{eq:S_w} becomes $S'_{W}=-\int^1_0d\tau(p_{\mu}+qA_{\mu})\dot{x}^{\mu}=-\int^1_0d\tau p_{\mu}\dot{x}^{\mu}/2= -Tp^2/(2m^*)$ based on $p^2$ is constant.

Finally using Eqs.~\eqref{eq:T_u_+}, ~\eqref{eq:mass_shell},  and~\eqref{eq:augmented_mass}, we can find that the action for the worldline instanton for the $u_+$ solution is 
\begin{equation}
S_{u_+,n}=\frac{n\pi i(m-\omega_{5})^{2}}{q\lambda_{E}}\,.
\end{equation}
The threshold with which the Schwinger mechanism is to overcome has been effectively reduced in mass by the axial gauge. However, we still must evaluate the remaining three other solutions.

Next, we evaluate instanton solutions associated with $v_+$.  However, notice that the mass-shell constraint leading to Eq.~\eqref{eq:+_constraint} is the same as it was for $u_+$. This is not all. In fact, for $v_+$, too, one can confirm in analogy to Eq.~\eqref{eq:u_+_velocity} that
\begin{equation}
\bar{v}_{+}\gamma_{\mu}v_{+}=\frac{1}{m^{*}}p_{\mu}=\frac{1}{T}\dot{x}_{\mu}\,,
\end{equation}
which is the same as for the $u_+$ case. Furthermore, we find all the equations of motion and constraints are identical to the $u_+$ case; thus we are led to
\begin{equation}
S_{v_+,n}=S_{u_+,n}\,,
\end{equation}
since the Berry phases vanish for both.  

Let us now last evaluate solutions associated with $u_-$ and $v_-$.  In analogy to $u_+$ and $v_+$ we find here too that the solutions for either $u_-$ or $v_-$ are identical to the other, therefore let us just treat $u_-$.  The mass-shell constraint here changes according to Eq.~\eqref{eq:-_constraint}.  However, because of the constraint, for $\bar{u}_-\gamma_\mu u_-=p_\mu/m^*$ we find that the definition of the effective mass is the same as the one given in Eq.~\eqref{eq:eff_mass}; what we find will change is the momentum invariant.  Recall that two solutions are possible for the momentum invariant: $p^2=(m\pm \omega_5)^2$. However,  here after inserting both into Eq.~\eqref{eq:-_constraint} we now find that
\begin{equation}
p^2=(m+\omega_5)^2\,,
\end{equation}
c.f., Eq.~\eqref{eq:mass_shell},  and hence also
\begin{equation}\label{eq:m_+}
m^*=m+\omega_5\,,
\end{equation}
c.f., Eq.~\eqref{eq:augmented_mass}.  Last, here our stationary points are now at Eq.~\eqref{eq:T_u_+} with Eq.~\eqref{eq:m_+} above.  Again, our equations of motion are identical to before with the only change being the mass-shell constraint leading to the above augmented mass.  Therefore,  one can confirm in this case that
\begin{equation}
S_{u_-,n}=S_{v_-,n}=\frac{n\pi i(m+\omega_5)^2}{q\lambda_E}\,.
\end{equation}

Gathering all the solutions one can finally find for the worldline instanton evaluated imaginary part of the effective action 
\begin{equation}
\textrm{Im}\Gamma(\lambda_E,\omega_{5})\approx2\sum_{n=1}^\infty\Bigl\{ e^{-\frac{n\pi}{q\lambda_{E}}(m-\omega_{5})^{2}}+e^{-\frac{n\pi}{q\lambda_{E}}(m+\omega_{5})^{2}}\Bigr\}\,.
\end{equation}
The most important observation in the above is that the exponential quadratic negative mass shift, found before in Eq.~\eqref{eq:res_final}, persists in the semiclassical approach. Therefore we can see the pair production enhancement from a spatial axial gauge, or vorticity, is a robust feature. 
The origin of the mass shift can be traced back to the augmented Dirac equation (\ref{eq:Dirac_electric}). The spatial axial gauge term can be rearranged as an energy shift by the spin-vorticity coupling $\gamma^{0} (D^0 + i \gamma^{2}\gamma^3 \omega_{5})$, where $i \gamma^{2}\gamma^3$ is the spin operator along $\omega_5$ applied there. 
Even though there is a decreased probability of pair production for the $u_-$ and $v_-$ cases, the overall pair production is favorably enhanced due to the $u_+$ and $v_+$ cases.

Let us finally mention that the results here agree--or rather are proportional to, since we are neglecting a prefactor contribution here--with those found in the previous section for $n=1$ to $\mathcal{O}(\omega_5^2)$, or alternatively in the $m\gg\omega_{5}$ limit. The discrepancy at higher orders may be attributed to fluctuations about the instantons, and or from contributions of off-diagonal terms in the Berry phase. Nevertheless we most importantly see the overall subtraction of the mass in the exponential due to the spatial axial gauge.

\section{Massless effective action with axial gauge}
\label{sec:mass}

Above we analyzed the one-loop effective action with massive fields, and alternatively one could arrive at corresponding observables for the nearly massless case by carefully taking the massless limit.  The effective action determined from a purely massless case, i.e., a theory which begins with no mass term, in fact need not be the same. Namely, even in the case of no axial gauge, i.e., 
\begin{align}
  & \lim_{m\rightarrow0}\int\mathcal{D}\bar{\psi}\mathcal{D}\psi\,e^{i\int d^{4}x\bar{\psi}[i\slashed{D}-m]\psi}\nonumber \\
&\neq\int\mathcal{D}\bar{\psi}\mathcal{D}\psi\,e^{i\int d^{4}x\bar{\psi}i\slashed{D}\psi}\eqqcolon e^{i\Gamma_{\textrm{massless}}[A,0]}\,.\label{eq:mass_massless}
\end{align}
With the axial gauge, clarifying the differences between the two effective actions is important in that we will find markedly different physics emerge. Notably, the massless case possesses an exact and simple solution through an anomalous phase in the fermion determinant via the Fujikawa method~\cite{PhysRevLett.42.1195,*PhysRevD.21.2848,*PhysRevD.22.1499}, whereas in the massless limit case no such term is present (for homogeneous fields). First let us examine the massive case perturbatively in $A_5^\mu$,  as doing so provides context for the above perturbative results.

Let us begin by writing down the massive effective action, c.f., Eq.~\eqref{eq:eff_action_def} for arbitrary $A_5^\mu$ as
\begin{equation}\label{eq:eff_action_def2}
e^{i\Gamma[A,A_{5}]}=\int\mathcal{D}\bar{\psi}\mathcal{D}\psi e^{i\int d^{4}x\bar{\psi}[\hat{\slashed{\Pi}}_5-m]\psi}\,.
\end{equation}
Then to define a perturbative expansion about $A_{5}$, let us look at a formal functional expansion \cite{Peskin:257493} of the fermionic determinant that leads to
\begin{equation}\label{eq:pert_def}
\Gamma[A,A_{5}]=\Gamma[A,0]+\sum_{n=1}^{\infty}\frac{i}{n}\textrm{Tr}\Bigl[\frac{-1}{i\slashed{D}-m}\gamma_{5}\slashed{A_{5}}\Bigr]^{n}\,.
\end{equation}
One can then find that to $\mathcal{O}(A_{5}^{3})$
\begin{align}
\Gamma&[A,A_{5}]\approx \Gamma[A,0]+i\int d^{4}x\lim_{x\rightarrow y}\textrm{tr}[\gamma_{5}\slashed{A_{5}}S^{\textrm{c}}(x,y)]\nonumber \\
 &+\frac{i}{2}\int d^{4}xd^{4}y\,\textrm{tr}[\gamma_{5}\slashed{A_{5}}S^{\textrm{c}}(x,y)\gamma_{5}\slashed{A_{5}}S^{\textrm{c}}(y,x)\Bigr]\,,\label{eq:pert_1}
\end{align}
where the electromagnetic dressed propagator satisfies $(i\slashed{D}_{x}-m)S^{c}(x,y)=-\delta^{4}(x-y)$.  It can be seen that the effective action admits an expansion whereby
the dressed propagator interacts with any number of interactions of an external axial gauge field of $\gamma_{5}\slashed{A}_{5}$. 

For concreteness let us examine homogeneous electromagnetic fields; here $S^c$ is well-known~\cite{PhysRev.82.664}. Then for the point split in-out propagator in the coincidence limit we take $\lim_{x\rightarrow y}S^{\textrm{c}}(x,y)\coloneqq\lim_{\epsilon\rightarrow0}(1/2)[S^{\textrm{c}}(x,x+\epsilon)+S^{\textrm{c}}(x+\epsilon,x)]$, which leads to a vanishing axial current, i.e.,  $j_{5}^{\mu}=\textrm{tr}\lim_{x\rightarrow y}\gamma^{\mu}\gamma_{5}S^{\textrm{c}}(x,y)=0$. See Ref.~\cite{PhysRevLett.121.261602,*doi:10.1142/S0217751X2030015X} for details.  Therefore the $\mathcal{O}(A_{5})$ term in the massive effective action, Eq.~\eqref{eq:pert_1}, vanishes. Higher order terms,  as evidenced above,  do not vanish. And importantly, for the purely massless case we will show that only the $\mathcal{O}(A_{5})$ term is present.  

Using the Fujikawa method~\cite{PhysRevLett.42.1195,*PhysRevD.21.2848,*PhysRevD.22.1499} one may determine an entirely different perturbative scheme about the axial gauge.  Let us take an axial gauge rotation in Eq.~\eqref{eq:eff_action_def2},  with $\psi\rightarrow\exp(-i\gamma_{5}A_{5\mu}x^{\mu})\psi$, and then 
\begin{align}
e^{i\Gamma[A,A_{5}]}&=\int\mathcal{D}\bar{\psi}\mathcal{D}\psi\exp(2i\textrm{Tr}\gamma_{5}A_{5\mu}x^{\mu}) \nonumber\\
&\times \exp\Bigl\{i\int d^{4}x\bar{\psi}[i\slashed{D}-me^{-2i\gamma_{5}A_{5}^{\mu}x_{\mu}}]\psi\Bigr\}\,,
\end{align}
where we remind the reader that the functional trace of $\gamma_5$ requires regulation and is in general non-vanishing, this is more easily recognizable with angle $\theta(x)=A_{5\mu}x^{\mu}$. The functional trace term, $\textrm{Tr}\gamma_{5}A_{5\mu}x^{\mu}$,
leads to the anomalous Chern-Simons term in QED. Let us perform the perturbative expansion as was done before for Eq.~\eqref{eq:pert_def}. One can eventually find to $\mathcal{O}(A_{5}^{3})$ that
\begin{align}
\Gamma&[A,A_{5}]=\Gamma[A,0]+2\textrm{Tr}\gamma_{5}A_{5\mu}x^{\mu}\nonumber \\
&+2im\int d^{4}x\,\textrm{tr}[\{i\gamma_{5}A_{5}^{\mu}x_{\mu}+(A_{5}^{\mu}x_{\mu})^{2}\}\lim_{x\rightarrow y}S^{\textrm{c}}(x,y)]\nonumber \\
 &-2im^{2}\int d^{4}xd^{4}y\,\textrm{tr}[S^{\textrm{c}}(x,y)\gamma_{5}A_{5}^{\mu}y_{\mu}S^{\textrm{c}}(y,x)\gamma_{5}A_{5}^{\mu}x_{\mu}]\,.\label{eq:pert_2}
\end{align}
Here we have acquired a new perturbative expansion about $A_{5}$ using the Fujikawa method.  At $\mathcal{O}(A_{5})$ we can then see
that we have a statement of the axial-Ward identity considering Eq.~\eqref{eq:pert_1}.  Therefore we can understand for massive fermions in homogeneous fields why the $\mathcal{O}(A_{5})$ term vanishes as a consequence of the vanishing axial-Ward identity~\cite{PhysRevLett.121.261602,*doi:10.1142/S0217751X2030015X}.  Yet, no such vanishing occurs for the purely massless case.

The purely massless effective action for an arbitrary field can be read off of Eq.~\eqref{eq:pert_2},  and is
\begin{equation}
\Gamma_{\textrm{massless}}[A,A_{5}]=\Gamma_{\textrm{massless}}[A,0]+2\textrm{Tr}\gamma_{5}A_{5\mu}x^{\mu}\,.
\end{equation}
The full massless effective action with an axial gauge potential is augmented only through the anomalous term arising from the fermion determinant.  Let us next evaluate the functional trace for an arbitrary field.  We define the regularized trace through eigenmodes of the quadratic form, $\hat{\slashed{\Pi}}_5$.  Even though our starting Lagrangian is augmented with the axial gauge,  its zero modes are the same as the case without an axial gauge since the axial gauge can be removed by a redefinition,  and therefore it is expected that the index theorem~\cite{ATIYAH1978185,*PhysRevD.14.3432} for both QED and QED with an axial gauge be the same.  Let us demonstrate this identification to $\mathcal{O}(A_{5}^{2})$.  To begin we define for $\Lambda\rightarrow\infty$  with 
\begin{equation}
I_\textrm{CS}\equiv \textrm{Tr}\theta \gamma_5 =\int d^{4}x\,\theta\, \textrm{tr}\langle x|\gamma_{5}\frac{\Lambda^{2}}{-\hat{\slashed{\Pi}}_5^2+\Lambda^{2}}|x\rangle\,,
\end{equation}
which agrees with a conventional quadratic in operator regularization if no axial gauge field were present \cite{Vandoren:2008xg}. Let us cast the functional trace in Schwinger propertime (where we have performed a propertime redefinition) as 
\begin{align}
I_\textrm{CS}& =\int d^{4}x\,\theta\,\textrm{tr}\int_{0}^{\infty}dT\,i\gamma_{5}e^{-iT}\langle x|e^{\frac{i}{\Lambda^{2}}\hat{\slashed{\Pi}}_5^2T}|x\rangle\,.
\end{align}
One may express the quadratic operator as \cite{Hattori_review} 
\begin{equation}\label{eq:quad_form}
-\hat{\slashed{\Pi}}_5^2=\slashed{D}^{2}+A_{5}^2+i\gamma_{5}[D_{\mu}A_{5}^{\mu}+2i\sigma_{\mu\nu}A_{5}^{\mu}D^{\nu}]\,. 
\end{equation} 
Then expanding to $\mathcal{O}(A_{5}^{2})$ and treating only the lowest order terms in the electromagnetic coupling that are non-vanishing after taking the trace we find that
\begin{equation}\label{eq:quad_operator}
I_\textrm{CS}=\int d^{4}x\,\theta\,\textrm{tr}\langle x|\gamma_{5}\frac{\Lambda^{2}}{\slashed{D}^{2}+\Lambda^{2}}|x\rangle\,.
\end{equation}
Therefore we can see that the functional trace is the same as the one in QED and hence familiar manipulations lead to \cite{Peskin:257493} 
\begin{equation}
I_\textrm{CS}=-\int d^{4}x\,\theta\,\frac{q^{2}}{8\pi^{2}}\widetilde{F}_{\mu\nu}F^{\mu\nu}\,.
\end{equation}
One can see that if we were to have homogeneous fields such a term would trivially vanish due to the linear $x$ under the integrand.  However, this does not entail that new physics does not emerge for the purely massless case.  Indeed one may calculate the vector current: $j_{\mu}=2 \delta I_\textrm{CS}/\delta A^{\mu}$ to find the chiral magnetic effect~\cite{Vilenkin:1980fu,*Nielsen:1983rb,*Alekseev:1998ds,*Fukushima:2008xe} for $A_{5}^{0}=\mu_{5}$:
\begin{equation}
\textbf{j}=\frac{q^{2}}{2\pi^{2}}\mu_{5}\textbf{B}\,,
\end{equation}
and also for $\boldsymbol{A}_{5}=\boldsymbol{\omega}_5$, the vorticity, one can find corrections to the charge density \cite{PhysRevLett.117.152002} such that 
\begin{equation}
j^0=\frac{q^{2}}{2\pi^{2}}\boldsymbol{\omega}_5\cdot\boldsymbol{B}\,.
\end{equation}
However such terms would not be present for the massive case in the massless limit according to Eq.~\eqref{eq:pert_2} since the Chern Simons term is matched by the pseudoscalar condensate term via the axial-Ward identity.

\section{Conclusions}
\label{sec:conclusions}

Augmentations to the one-loop Euler-Heisenberg Lagrangian with an axial gauge coupling have been examined. For the massive case (as well as massless limit of a massive theory) we confined our attention to two configurations with an exact eigendecomposition: 1.~a magnetic field with chiral chemical potential and 2.~an electric field with spatial axial gauge field (that has been argued to resemble a vorticity).  
In a perturbative expansion about $A_5^\mu$ for weak fields in $\textrm{Re}[\mathbb{L}_\text{EH}]$, it was determined in the absence of electromagnetic fields that an overall increase (decrease) of the system energy was present for the chiral chemical potential (spatial axial gauge).

An important feature was observed in $\textrm{Im}[\mathbb{L}_\text{EH}]$ for the case of the electric field with a spatial axial gauge in that the spatial axial gauge enhanced the Schwinger pair production. This was shown perturbatively in $\mathcal{O}(A_5)$ through the eigendecomposition approach; an all-orders expression was also derived. Using a semi-classical  worldline instanton approach it was determined that the enhancement was possible from a negative mass shift. The enhancement occurs at the exponential level and is therefore thought significant in systems with vorticity.

The massless case (where no mass term is present in the partition function) had an exact solution thanks to the Fujikawa method. It was demonstrated that an axial gauge could be rotated away producing a Chern-Simons term as a corrective factor to the massless Euler-Heisenberg Lagrangian.

An important feature not discussed here is the massive case with simultaneous electric and magnetic fields such that $\boldsymbol{E}\cdot \boldsymbol{B}\neq 0$. Although the two configurations examined here have exact eigendecompositions, we cannot study the anomalous features present in the massless case. One should anticipate such features for the massive case, because they are present even without an axial gauge, leading to a description of the axial-Ward identity. Therefore, it is an important extension to this work to analyze the case with $\boldsymbol{E}\cdot \boldsymbol{B}\neq 0$ under an axial gauge field.

\begin{acknowledgments}
This work is partially supported by JSPS KAKENHI under grant Nos.~20K03948 and 22H02316 and Taiwan National Science and Technology Council (NSTC) under grant No. MOST 110-2112-M-001-070-MY3. 
\end{acknowledgments}

\appendix*
\section{Steepest Descents in $z$ and $\bar{z}$}
\label{sec:appendix}

It was argued in Sec.~\ref{sec:worldline} that worldline instantons can be found from evaluating the coupled equations given in Eqs.~\eqref{eq:eom_x}-\eqref{eq:eom_z}. However, rather than expanding about stationary points in $T$ and then evaluating Eq.~\eqref{eq:P_5} for $z$ and $\bar{z}$, one may also expand about stationary points in $z$ and $\bar{z}$, and then later evaluate the $T$ integral. Then in the latter case the equations of motion leading to worldline instantons would be
\begin{align}
    \dot{x}_{\mu}&=T\bar{z}\gamma_{\mu}z\,,\label{eq:eomx_app}\\
    \dot{p}_{\mu}&=qF_{\mu\nu}\dot{x}^{\nu}\,,\\
    \dot{\bar{z}}&=-iT\bar{z}\slashed{\Pi}_5\,\label{eq:zbar_app}\\
    \dot{z}&=iT\slashed{\Pi}_5z\label{eq:z_app}\,.
\end{align}
In Sec.~\ref{sec:worldline} we took the approach of expanding about $T$ primarily for two reasons:
\begin{enumerate}
    \item In the case without an axial gauge field it is known that the spinor degrees of freedom do not affect worldline instantons in homogeneous fields.
    \item Furthermore, we have found above (e.g. Eq.~\eqref{eq:Im_sum_poles}), that all poles in the exact expression can be located independent of the axial gauge field, and for that matter spin degrees of freedom. 
\end{enumerate}
Nevertheless, let us outline the alternative option of evaluating Eqs.~\eqref{eq:zbar_app}-\eqref{eq:z_app} here, and in so doing identify their incompatibility with the previous solution.
 
 One may find a general solution to Eq.~\eqref{eq:z_app} in the form similar to the Wilson loop encountered earlier in Eq.~\eqref{eq:path_int_before} as
 \begin{equation}\label{eq:zsoln_app}
     z(\tau)=\mathcal{P}\exp\Bigl[iT\int^\tau_0\slashed{\Pi}_5\Bigr]z(0)\,,
 \end{equation}
 and likewise for $\bar{z}$. Then here, the worldline action \textit{after} taking the steepest descents reads
 \begin{equation}
     S'_\text{W}=\int^1_0[-mT-p_\mu\dot{x}^\mu -qA_\mu \dot{x}^\mu]\,.
 \end{equation}
 where worldline instantons have $T$ dependence and are now solutions to Eqs.~\eqref{eq:eomx_app}-\eqref{eq:z_app}, which need not be the same as were discovered before. Now, an incompatibility can be seen if the previous mass-shell constraint is utilized, namely Eq.~\eqref{eq:P_5}, leading to the spinor-like solutions. For then, one would find only a propertime independent solution to Eq.~\eqref{eq:zsoln_app}, leading to a trivial solution considering the constraint given in Eq.~\eqref{eq:eom_z}.
 
 One caveat of taking the $T$ integral and evaluating for $z$ and $\bar{z}$ can be seen in that what would be off-diagonal terms are neglected in the geometric or Berry~\cite{doi:10.1098/rspa.1984.0023} phase term, e.g., $\bar{u}_i\dot{z}\sim 0$ for $z\neq u_i$. Therefore we can see that our approach is equivalent to the adiabatic theorem wherein each eigenvector persists and no level jumping between eigenvectors may occur. To see this connection, one need only diagonalize the path ordered expression in Eq.~\eqref{eq:path_int_before}, giving rise to a geometric phase whose diagonal entries then describe our current setup. It is important to ascertain what precisely the adiabatic approximation entails for our setup. The adiabatic theorem in the context of Berry's phase is typically employed for large times (here Schwinger propertimes) and level separation. For an analogous setup please see~\cite{PhysRevD.105.116014}. One may define the adiabatic theorem in the worldline setting as $2|\lambda| T \gg 2\pi$, for $\lambda$ an arbitrary gap later given by the eigenvalues in Eq.~\eqref{eq:m_eigenvalue}. This procedure is ordinarily marred before having taken the $T$ integral; however, here we may later treat this inequality with more rigor since the $T$ integral will be evaluated at specific points, albeit on the imaginary propertime axis. Since the adiabatic theorem must hold even after analytic continuation in propertime to the negative imaginary propertime axis (the Wick rotation is guaranteed because of the $i\epsilon$ and quadratic mass prescription), one may equally well state the adiabatic theorem as $ 2 i|\lambda| T \gg 2\pi$ for $T\rightarrow i T$. We discover through the worldline instanton method in Eq.~\eqref{eq:m_eigenvalue} that the gap will occur at the mass-shell, $|\lambda|=m$, and on the stationary points given in Eq.~\eqref{eq:T_u_+} for both $m^*$ leading to the following inequalities:
 \begin{equation}
     \frac{2n m(m\pm\omega_5)}{q\lambda_E}\gg 1\,,
 \end{equation}
 for all $n$ instanton. One may recognize the above inequality as the weak fields/large mass approximation well-known in the worldline instanton configuration space path integral~\cite{AFFLECK1982509,*PhysRevD.72.105004} for large mass (a factor of $\pi$ is, however, not present on the right hand side). Thus we can see that in the phase space path integral the adiabatic theorem ensures the weak field/large mass approximation.

\bibliography{references_reply}
\bibliographystyle{apsrev4-1}
\end{document}